\documentclass{elsarticle}

\usepackage[reviewcopy]{adndt}
\usepackage{longtable}

\usepackage{amsmath}

\usepackage{amssymb}

\biboptions{square,sort&compress}
\bibpunct[]{[}{]}{,}{n}{}{;}
\begin{document}

\begin{frontmatter}


\title{Discovery of Isotopes of the Transuranium Elements with 93 $\le$ Z $\le$ 98}


\author{C. Fry}
\author{M. Thoennessen\corref{cor1}}\ead{thoennessen@nscl.msu.edu}

 \cortext[cor1]{Corresponding author.}

 \address{National Superconducting Cyclotron Laboratory and \\ Department of Physics and Astronomy, Michigan State University, \\ East Lansing, MI 48824, USA}

\begin{abstract}
One hundred and five isotopes of the transuranium elements neptunium, plutonium, americium, curium, berkelium and californium have so far been observed; the discovery of these isotopes is discussed. For each isotope a brief summary of the first refereed publication, including the production and identification method, is presented.
\end{abstract}

\end{frontmatter}





\newpage
\tableofcontents
\listofDtables

\vskip5pc

\section{Introduction}\label{s:intro}

The discovery of neptunium, plutonium, americium, curium, berkelium and californium isotopes is discussed as part of the series summarizing the discovery of isotopes, beginning with the cerium isotopes in 2009 \cite{2009Gin01}. Guidelines for assigning credit for discovery are (1) clear identification, either through decay-curves and relationships to other known isotopes, particle or $\gamma$-ray spectra, or unique mass and Z-identification, and (2) publication of the discovery in a refereed journal. The authors and year of the first publication, the laboratory where the isotopes were produced as well as the production and identification methods are discussed. When appropriate, references to conference proceedings, internal reports, and theses are included. When a discovery includes a half-life measurement, the measured value is compared to the currently adopted value taken from the NUBASE evaluation \cite{2003Aud01} which is based on the ENSDF database \cite{2008ENS01}. In cases where the reported half-life differed significantly from the adopted half-life (up to approximately a factor of two), we searched the subsequent literature for indications that the measurement was erroneous. If that was not the case we credited the authors with the discovery in spite of the inaccurate half-life. All reported half-lives inconsistent with the presently adopted half-life for the ground state were compared to isomer half-lives and accepted as discoveries if appropriate following the criterium described above.

The first criterium is not clear cut and in many instances debatable. Within the scope of the present project it is not possible to scrutinize each paper for the accuracy of the experimental data as is done for the discovery of elements \cite{1991IUP01}. In some cases an initial tentative assignment is not specifically confirmed in later papers and the first assignment is tacitly accepted by the community. The readers are encouraged to contact the authors if they disagree with an assignment because they are aware of an earlier paper or if they found evidence that the data of the chosen paper were incorrect.

The discovery of several isotopes has only been reported in conference proceedings which are not accepted according to the second criterium. One example from fragmentation experiments why publications in conference proceedings should not be considered are $^{118}$Tc and $^{120}$Ru which had been reported as being discovered in a conference proceeding \cite{1996Cza01} but not in the subsequent refereed publication \cite{1997Ber01}.





In contrast to the criteria for the discovery of an element \cite{1976Har02,1990Sea01,1991IUP01}, the criteria for the discovery or even the existence of an isotope are not well defined (see for example the discussion in reference \cite{2004Tho01}). Therefore it is possible, for example in the case of plutonium and californium, that the discovery of an element does not necessarily coincide with the first discovery of a specific isotope.

The initial literature search was performed using the databases ENSDF \cite{2008ENS01} and NSR \cite{2008NSR01} of the National Nuclear Data Center at Brookhaven National Laboratory. These databases are complete and reliable back to the early 1960's. For earlier references, several editions of the Table of Isotopes were used \cite{1948Sea01,1953Hol02,1958Str01,1967Led01}. Additional excellent resources were the preface of the National Nuclear Energy Series volume containing the Transuranium research papers \cite{1949Sea04} and the book ``The elements beyond Uranium'' by Seaborg and Loveland \cite{1990Sea01}.



\section{Discovery of $^{225-244}$Np}\vspace{0.0cm}

The element neptunium was discovered by McMillan and Abelson in 1940 \cite{1940McM01}. A previous claim of natural occurring element 93 in 1934, named bohemium \cite{1934Kob01}, was very quickly discredited \cite{1934Spe01}. Also in 1934 Fermi et al.\ reported the ``Possible production of elements of atomic number higher than 92'' in neutron bombardment of uranium \cite{1934Fer02}. In his Nobel lecture he named the element with Z=93 ausenium but in the write-up of the lecture he added a footnote: ``The discovery by Hahn and Strassmann of barium among the disintegration products of bombarded uranium, as a consequence of a process in which uranium splits into two approximately equal parts, makes it necessary to reexamine all the problems of the transuranic elements, as many of them might be found to be products of a splitting of uranium.'' \cite{1938Fer01}. In 1939, another claim for natural occurring neptunium, named sequanium \cite{1939Hul01}, was later shown to be incorrect.

The name neptunium was officially accepted at the 15$^{th}$ IUPAC conference in Amsterdam in 1949 \cite{1949IUP01,2005Kop01}.

\begin{figure}
	\centering
    \includegraphics[scale=.7]{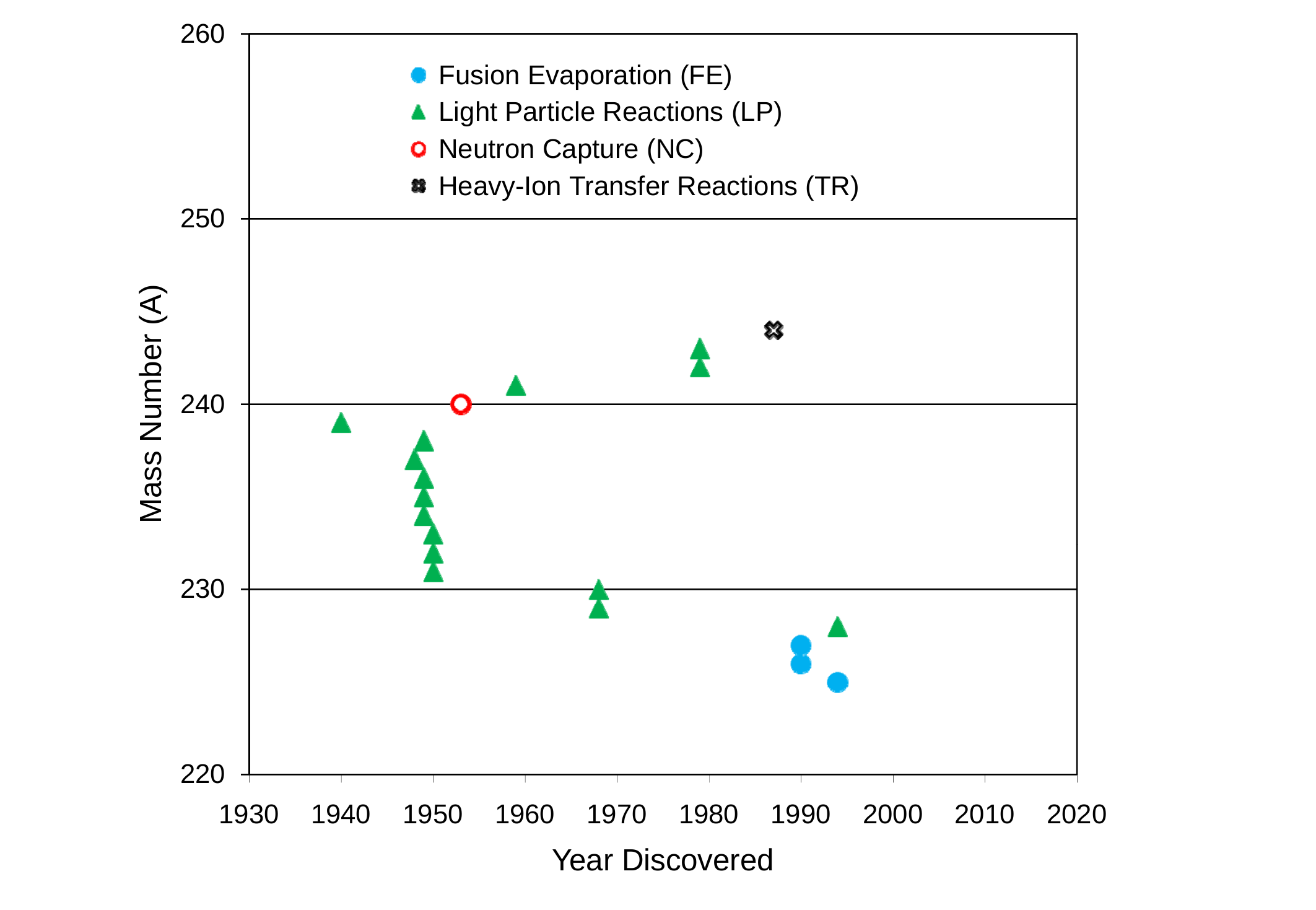}
	\caption{Neptunium isotopes as a function of time when they were discovered. The different production methods are indicated.}
\label{f:year-neptunium}
\end{figure}

Twenty neptunium isotopes from A = 225--244 have been discovered so far. According to the HFB-14 model \cite{2007Gor01} about 80 additional neptunium isotopes could exist. Figure \ref{f:year-neptunium} summarizes the year of first discovery for all neptunium isotopes identified by the method of discovery: heavy-ion (A$>$4) fusion evaporation reactions (FE), light-particle (A$\le$4) reactions (LP), neutron capture (NC), and heavy-ion transfer reactions (TR). In the following, the discovery of each neptunium isotope is discussed in detail and a summary is presented in Table 1.

\subsection*{$^{225}$Np}
The 1994 article ``The kinematic separator VASSILISSA performance and experimental results'' by Yeremin et al.\ was the first publication reporting the observation of $^{225}$Np \cite{1994Yer01}. $^{22}$Ne beams from the Dubna U-400 cyclotron bombarded a $^{209}$Bi target forming $^{225}$Np in the (6n) fusion-evaporation reaction. $^{225}$Np was separated with the kinematic separator VASSILISSA and implanted in an array of silicon detectors which also recorded subsequent $\alpha$ decays. ``New isotopes $^{218,219, 223-226}$U, $^{225-227}$Np and $^{228-230}$Pu were produced and identified, and their $\alpha$-decay energy and half life times were measured.'' An $\alpha$-energy of 8630(2)~keV for $^{225}$Np is listed in a table.

\subsection*{$^{226}$Np}
In 1990 Ninov et al.\ reported the discovery of $^{226}$Np in ``Identification of neutron-deficient isotopes $^{226,227}$Np'' \cite{1990Nin01}. A $^{209}$Bi target was bombarded with 5.5~MeV/u $^{22}$Ne beams from the GSI UNILAC accelerator populating $^{226}$Np in the (5n) fusion-evaporation reaction. Evaporation residues were separated with the gas-filled DQQ spectrometer NASE and implanted into a passivated ion implanted silicon detector which also recorded subsequent $\alpha$ decays. ``The spectrum shows two $\alpha$ lines at (8044$\pm$20)~keV and (8430$\pm$20)~keV which we assigned to $^{226}$Np and its granddaughter $^{218}$Fr.'' The measured half-life of 30(8)~ms is included in the calculation of the currently adopted value.

\subsection*{$^{227}$Np}
The 1990 discovery of $^{227}$Np by Andreyev et al.\ was reported in ``A new isotope and $\alpha$-lines in the Th-Np region and their production cross-sections'' \cite{1990And01}. Beams of 106$-$115~MeV $^{22}$Ne from the Dubna U-400 cyclotron bombarded a $^{209}$Bi target forming $^{227}$Np in the (4n) fusion-evaporation reaction. $^{227}$Np was separated with  the VASSILISSA kinematic separator and implanted in seven surface barrier detectors which also recorded subsequently emitted $\alpha$ particles. ``We explain these correlation groups as due to the $\alpha$-decay chain of the new isotope of $^{227}$Np. The time distribution of the events in the correlation groups observed at 8.00~MeV and 8.20~MeV supports this conclusion.'' In a table an $\alpha$ energy of 7680(10)~keV is listed for $^{227}$Np. Less than five months later Ninov et al.\ independently reported $\alpha$-decay energies of 7650(20)~keV and 7677(20)~keV with a 510(60)~ms half-life \cite{1990Nin01}.

\subsection*{$^{228}$Np}
Kreek et al.\ described the observation of $^{228}$Np in the 1994 paper ``Electron-capture delayed fission properties of $^{228}$Np'' \cite{1994Kre01}. A stack of 23 $^{233}$U foils were irradiated with a 50~MeV proton beam from the Berkeley 88-inch cyclotron forming $^{228}$Np in the (p,6n) reaction. Electron-capture delayed fission was measured following chemical separation. ``ECDF was studied in $^{228}$Np produced via the $^{233}$U(p,6n)$^{228}$Np reaction. The fission properties and half-life were measured with a rotating-wheel system. The half-life of this isotope was determined to be 61.4$\pm$1.4~s from measurements of the fission activity.'' This value is the currently adopted half-life. Previously, Kuznetsov et al.\ had assigned a 60~sec half-life to either $^{227}$Np or $^{228}$Np \cite{1967Kuz05}.

\subsection*{$^{229,230}$Np}
In the 1968 article ``New neptunium isotopes, $^{230}$Np and $^{229}$Np'' Hahn et al.\ reported the discovery of $^{229}$Np and $^{230}$Np \cite{1968Hah01}. Enriched $^{233}$U targets were bombarded with 32$-$41.6~MeV protons from the Oak Ridge Isochronous Cyclotron forming $^{229}$Np and $^{230}$Np in (p,5n) and (p,4n) reactions, respectively. Reaction products were implanted on a catcher foil which was periodically rotated in front of a surface barrier Si(Au) detector. ``From [the equation], it is apparent that the short-lived daughters attained transient equilibrium with the 4.0~min parent before counting began so no growth portion is seen in the data. The 6.89~MeV $\alpha$-particle is accordingly assigned to $^{229}$Np, the $\alpha$-decay progenitor of $^{225}$Pa... The $^{226}$Pa chain is clearly seen to grow in with a 1.8~min half-life and reach transient equilibrium with a parent of 4.6~min half-life. The 6.66~MeV $\alpha$ particle is thus assigned to the $\alpha$-precursor of $^{226}$Pa, namely $^{230}$Np.'' These half-lives of 4.0(2)~min and 4.6(3)~min are the presently adopted values for $^{229}$Np and $^{230}$Np, respectively.

\subsection*{$^{231-233}$Np}
In 1950, Magnusson et al.\ discovered $^{231}$Np, $^{232}$Np, and $^{233}$Np as described in the paper ``New isotopes of neptunium'' \cite{1950Mag01}. $^{231}$Np was produced by bombarding natural uranium with 100~MeV deuterons from the Berkeley 184-inch cyclotron. $^{232}$Np and $^{233}$Np were populated with 15~MeV deuterons from the 60-inch cyclotron on $^{233}$U targets. Decay and absorption curves were measured with a Geiger counter following chemical separation. In addition, $\alpha$-particle spectra were recorded with an argon-filled ionization chamber. ``The isotope Np$^{231}$ has a half-life of 50$\pm$3~min.\ and emits alpha-particles of 6.28-Mev energy; the proportion of decay by electron-capture has not been determined. The isotope Np$^{233}$ decays predominantly by electron-capture with a 35$\pm$3-min.\ half-life; it has an alpha-decay half-life roughly determined to be ca.\ 10 yr.\ corresponding to a K/$\alpha$-branching ratio of 1.5$\times$10$^5$ and the alpha-particles have an energy of 5.53~Mev. A 13$\pm$3-min.\ period with electromagnetic radiation, indicating orbital electron-capture, is attributed to Np$^{232}$.'' These half-lives of 50(3)~min for $^{231}$Np, 13(3)~min for $^{232}$Np and 35(3)~min for $^{233}$Np are close to the currently accepted values of 48.8(2)~min, 14.7(3)~min, and 36.2(1)~min, respectively.

\subsection*{$^{234}$Np}
The observation of $^{234}$Np was described in the 1949 paper ``Products of the deuteron and helium-ion bombardments of U$^{233}$'' by Hyde et al.\ \cite{1949Hyd01}. $^{233}$U targets were bombarded with 44~MeV $\alpha$ particles from the Berkeley 60-in.\ cyclotron. Alpha-particle absorption and decay spectra were measured following chemical separation. In addition, $\gamma$ rays were detected with an argon alcohol filled end-window tube. ``Np$^{234}$ is shown to be an orbital-electron-capturing isotope of 4.40$\pm$0.05 days half life with an associated $\gamma$ ray of 1.9 mev energy.'' This value agrees with the currently adopted half-life of 4.4(1)~d. James et al.\ had assigned a 4.5~d half-life to either $^{233}$Np or $^{234}$Np \cite{1949Jam01}.

\subsection*{$^{235-236}$Np}
James et al.\ identified $^{235}$Np and $^{236}$Np in  ``Products of helium-ion and deuteron bombardment of U$^{235}$ and U$^{238}$'' \cite{1949Jam01}. Natural uranium and $^{235}$U targets were bombarded with 16~MeV deuterons and 32~MeV $\alpha$ particles from the Berkeley 60-inch cyclotron. X rays, $\gamma$ rays and $\alpha$ particles were measured following chemical separation. ``Np$^{236}$ is a $\beta$-particle emitter with a half life of 22~hr. Its daughter, Pu$^{236}$, emits $\alpha$ particles with a range of 4.3~cm (energy 5.7~mev); it decays with a half life of 2.7 years. Np$^{235}$ decays by orbital-electron capture with a half life of approximately 400$\pm$20 days.'' These half-lives of 400(20)~d and 22~h agree with the presently accepted values of 396.1(12)~d for the ground state of $^{235}$Np and 22.5(4)~h for the isomeric state of $^{236}$Np, respectively.  The ground state of $^{236}$Np was first observed six years later \cite{1955Stu01}.

\subsection*{$^{237}$Np}
$^{237}$Np was identified in 1948 by Wahl and Seaborg in ``Nuclear properties of 93$^{237}$'' \cite{1948Wah01}. A uranyl nitrate hexahydrate target was irradiated with fast neutrons produced by bombarding a beryllium target with deuterons from the Berkeley 60-inch cyclotron forming $^{237}$U in the reaction $^{238}$U(n,2n). $^{237}$Np was then populated with $\beta$ decay. Resulting activities were measured with a Lauritsen electroscope following chemical separation. ``The alpha-counting rate of this sample, as mentioned above, was about 300 counts per minute, and from this value, together with the calibrated efficiency (45 percent) of the ionization chamber, it is calculated that the half-life of alpha-emitting 93$^{237}$ is about 3$\times$10$^6$~years.'' This value is close to the currently adopted half-life of 2.144$\times$10$^6$~y. The actual experiment was performed about seven years earlier: ``This article was mailed, as a secret report, from Berkeley, California to the Uranium Committee in Washington, D.\ C.\ on April 14, 1942. The experimental work was done during 1941 and the early part of 1942.''

\subsection*{$^{238}$Np}
Kennedy et al.\ published the observation $^{238}$Np in the 1949 paper ``Formation of the 50-year element 94 from deuteron bombardment of U$^{238}$'' \cite{1949Ken01}. An enriched $^{238}$U sample was bombarded with 14~MeV deuterons from the Berkeley 60-in.\ cyclotron. Following chemical separation, the growth of $\alpha$-particles was followed and aluminum absorption curves were taken on a Lauritsen electroscope. ``For these reasons, the most probable assignments of the 2.0-day element 93 and 50-year element 94 are 93$^{238}$ and 94$^{238}$ produced by the reactions U$^{238}$(d,2n)93$^{238}$, 93$^{238}\stackrel{\beta^-}{\longrightarrow}$ 2.0~day 94$^{238}$.'' This value for $^{238}$Np agrees with the currently adopted half-life of 2.117(2)~d. Earlier a 2~day activity was assigned to either $^{238}$Np, $^{236}$Np, or $^{235}$Np \cite{1946Sea01}.

\subsection*{$^{239}$Np}
In 1940 McMillan and Abelson discovered $^{239}$Np as reported in the paper ``Radioactive element 93'' \cite{1940McM01}. Uranium samples were activated with neutrons produced by the Berkeley cyclotron. Beta-decay curves were measured following chemical separation. ``This fact, together with the apparent similarity to uranium suggests that there may be a second `rare earth' group of similar elements starting with uranium. The final proof that the 2.3-day substance is the daughter of the 23-minute uranium is the demonstration of its growth from the latter...'' The figure caption of the decay curve states: ``Growth of 2.3-day 93$^{239}$ from 23-minute U$^{239}$.'' The quoted value agrees with the currently adopted half-life of 2.356(3)~d.

\subsection*{$^{240}$Np}
In 1950, the observation of $^{240}$Np was announced in the article ``The radiations of U$^{240}$ and Np$^{240}$'' by Knight et al.\ \cite{1953Kni01}. Uranium was irradiated with neutrons to form $^{240}$U which populated $^{240}$Np by $\beta$ decay. Following chemical separation, decay curves were measured with continuous-flow methane gas proportional counters and $\beta$- and $\gamma$-ray spectra were recorded with a magnetic lens spectrometer and a NaI(Tl) crystal, respectively. ``These measurements yielded a U$^{240}$ half-life of 14.1$\pm$0.2~hours, and a Np$^{240}$ half-life of 7.3$\pm$0.3~minutes.'' The quoted half-life corresponds to an isomeric state. Knight et al.\ credited Hyde and Studier with the discovery of $^{240}$Np quoting an unpublished report \cite{1948Hyd01}.

\subsection*{$^{241}$Np}
In the 1959 paper ``Decay of Np$^{241}$'' Vandenbosch identified $^{241}$Np \cite{1959Van01}. Uranium foils were irradiated with 32 and 43~MeV $\alpha$ particles from the Argonne cyclotron to form $^{241}$Np in the reaction $^{238}$U($\alpha$,p). Decay curves were recorded with 2$\pi$ and end-window proportional counters, and $\beta$- and $\gamma$-ray spectra were measured with an anthracene and sodium iodide crystal, respectively, following chemical separation. ``The decay of a 16-minute neptunium activity attributed to Np$^{241}$ has been studied with anthracene and sodium iodide scintillation counters. The principal mode of decay appears to be a beta group decaying to the ground state of Pu$^{241}$ with a beta end-point energy of 1.36$\pm$0.10~Mev.'' This value is close to the currently adopted half-life of 13.9(2)~min. The 1953 Table of Isotopes assigned a 60~min half-life to $^{241}$Np \cite{1953Hol02} based on an unpublished report \cite{1951Ort02}. This half-life was later reassigned to an isomer of $^{240}$Np \cite{1960Les01}.

\subsection*{$^{242}$Np}
Haustein et al.\ reported the observation of $^{242}$Np in the 1979 paper ``Identification and decay of $^{242}$U and $^{242}$Np'' \cite{1979Hau01}. $^{244}$Pu targets were irradiated with 30$-$160~MeV neutrons produced at the Brookhaven Medium Energy Intense Neutron (MEIN) facility by bombarding a water-cooled copper beam stop with 200~MeV protons from the Alternating Gradient Synchrotron. Gamma- and beta-rays were measured with Ge(Li) and plastic detectors, respectively, following chemical separation. ``By combining the data from several of the most intense runs we have by least square analyses T$_{1/2}$=16.8$\pm$0.5~min for $^{242}$U and T$_{1/2}$=2.2$\pm$0.2~min for $^{242}$Np'' This half-life is the currently adopted value for $^{242}$Np.

\subsection*{$^{243}$Np}
In the 1979 paper ``(t,$\alpha$) reaction on actinide nuclei and the observation of $^{243}$Np'' Flynn et al.\ identified $^{243}$Np \cite{1979Fly01}. A $^{244}$Pu target was bombarded with 17~MeV tritons from the Los Alamos FN Van de Graaff facility populating $^{243}$Np in (t,$\alpha$) reactions. $^{243}$Np was identified by measuring the ejectiles with a quadrupole-dipole-dipole-dipole spectrometer. ``An extrapolation of ground state masses of the lighter neptuniums would suggest a Q value of 12.5$\pm$0.1. The observed Q value of the highest energy alpha group was 12.405$\pm$0.010~MeV, which corresponds to a $^{243}$Np mass of 243.064330~u for this state.'' The first half-life measurement was reported eight years later by Moody et al.\ \cite{1987Moo01} claiming discovery of $^{243}$Np without quoting the work by Flynn et al.

\subsection*{$^{244}$Np}
Moody et al.\ reported the discovery of $^{244}$Np in the 1987 paper ``New nuclides: neptunium-243 and neptunium-244'' \cite{1987Moo01}. A 1430~MeV $^{136}$Xe beam from the GSI UNILAC accelerator bombarded $^{244}$Pu targets. Gamma-ray spectra were measured with two germanium detectors following chemical separation. ``The decay of 2.29-min $^{244}$Np (probable J$^\pi$=7$^-$) populates the high-spin members of the ground state rotational band in $^{244}$Pu.'' This value is the currently adopted half-life.

\section{Discovery of $^{228-247}$Pu}\vspace{0.0cm}

The element plutonium was discovered by Seaborg et al.\ in December 1940 and submitted for publication in Physical Review on January 28, 1941. However, it was only published in 1946 as explained in a footnote of the paper: ``This letter was received for publication on the date indicated but was voluntarily withheld from publication until the end of the war'' \cite{1946Sea01}. This is an example where the element discovery differs from the discovery of a specific isotopes because no mass identification was made. The parent $\beta$-decay activity was assigned to either mass 238, 236 or 235 \cite{1946Sea01}.

In 1934 Fermi et al.\ had reported the ``Possible production of elements of atomic number higher than 92'' in neutron bombardment of uranium \cite{1934Fer02}. In his Nobel lecture he named the element with Z=94 hesperium, but in the write-up of the lecture he added a footnote: ``The discovery by Hahn and Strassmann of barium among the disintegration products of bombarded uranium, as a consequence of a process in which uranium splits into two approximately equal parts, makes it necessary to reexamine all the problems of the transuranic elements, as many of them might be found to be products of a splitting of uranium.'' \cite{1938Fer01}.

The name plutonium was officially accepted at the 15$^{th}$ IUPAC conference in Amsterdam in 1949 \cite{1949IUP01,2005Kop01}.

\begin{figure}
	\centering
    \includegraphics[scale=.7]{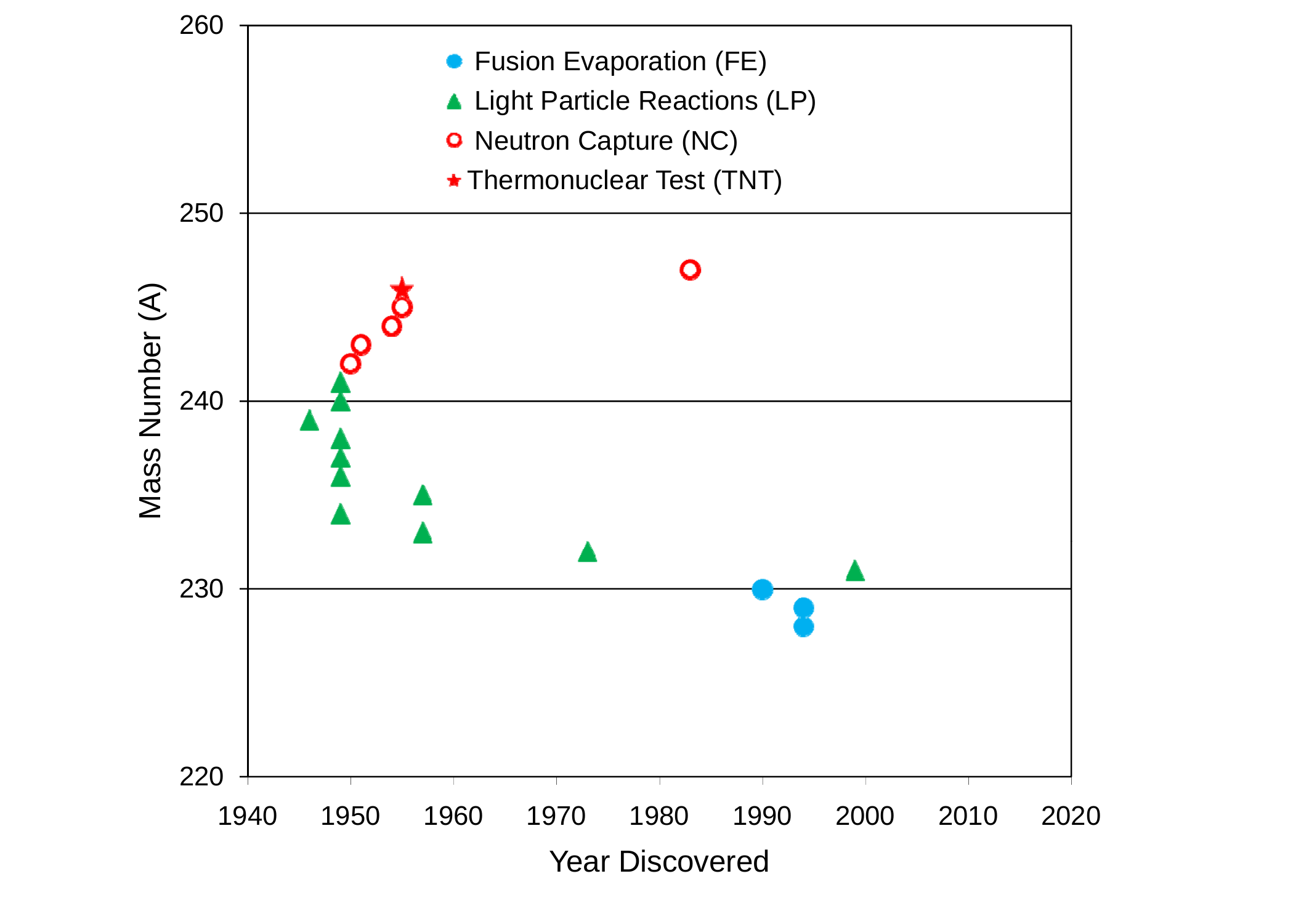}
	\caption{Plutonium isotopes as a function of time when they were discovered. The different production methods are indicated.}
\label{f:year-plutonium}
\end{figure}

Twenty plutonium isotopes from A = 228--247 have been discovered so far.  According to the HFB-14 model \cite{2007Gor01} about 85 additional plutonium isotopes could exist. Figure \ref{f:year-plutonium} summarizes the year of first discovery for all plutonium isotopes identified by the method of discovery: heavy-ion (A$>$4) fusion evaporation reactions (FE), light-particle (A$\le$4) reactions (LP), neutron capture (NC), and thermonuclear tests (TNT). In the following, the discovery of each plutonium isotope is discussed in detail and a summary is presented in Table 1.

\subsection*{$^{228,229}$Pu}
In 1994 Andreyev et al.\ reported the discovery of $^{228}$Pu and $^{229}$Pu in the paper ``New nuclides $^{228,229}$Pu'' \cite{1994And01}. Enriched $^{207}$Pb and $^{208}$Pb targets were bombarded with a 5.50~MeV/u $^{24}$Mg and a 5.58~MeV/u $^{26}$Mg beam from the Dubna U400 cyclotron. $^{228}$Pu was formed in the fusion-evaporation reaction $^{208}$Pb($^{24}$Mg,4n) and $^{229}$Pu was formed in the reactions $^{207}$Pb($^{26}$Mg,4n) and $^{208}$Pb($^{26}$Mg,5n). $^{228}$Pu and $^{229}$Pu were separated with the VASSILISSA electrostatic separator and implanted into a position sensitive silicon strip detector which also recorded subsequent $\alpha$ decay. ``Applying the time window of 0-4~ms three additional correlation chains starting from the $\alpha$-decays with the average energy of E$_{\alpha_I}$=(7810$\pm$20)~keV were found in the products of the $^{208}$Pb($^{24}$Mg,4n)$^{228}$Pu reaction at the beam energy of E/A=5.50~MeV/u. They were assigned to $^{228}$Pu on the basis of the genetic correlations with the $\alpha$ decays of known isotope $^{224}$U (E$_{\alpha_{II}}$=8470~keV) and the full sum of the pulses from its daughter products $^{224}$Th+$^{216}$Ra (E$_{\alpha_{III+IV}}$=18140~keV)... The assignment of the new isotope $^{229}$Pu was based on six correlations found, which are summarized in the Table.''

\subsection*{$^{230}$Pu}
Andreyev et al.\ discovered $^{230}$Pu as described in the 1990 paper ``New nuclide $^{230}$Pu'' \cite{1990And03}. An enriched $^{208}$Pb target was bombarded with a 135~MeV $^{26}$Mg beam from the Dubna U-400 cyclotron forming $^{230}$Pu in the (4n) fusion-evaporation reaction. $^{230}$Pu was separated with the kinematic separator VASSILISSA  and implanted into a silicon surface barrier detector which also recorded subsequent $\alpha$ decay. ``The new isotope $^{230}$Pu was identified according to the $\alpha$-$\alpha$ correlation to decays of its daughter nuclei $^{226}$U, $^{222}$Th and ($^{218}$Ra$+^{214}$Rn), see [the figure].''

\subsection*{$^{231}$Pu}
In the 1999 paper ``New plutonium isotope: $^{231}$Pu'' Laue et al.\ reported the first observation of $^{231}$Pu \cite{1999Lau01}. A stack of eleven $^{233}$U targets was irradiated with a 47.1~MeV $^3$He beam from the Berkeley 88-in.\ cyclotron. Alpha-particle correlations were measured with thirteen passivated ion implanted silicon (PIPS) detectors following chemical separation. ``$^{231}$Pu was positively identified. The half-life was determined to be (8.6$\pm$0.5)~min from the analysis of the $\alpha$-$\alpha$-correlations of the $^{223}$Th, $^{219}$Ra, and $^{215}$Rn daughters of its $\alpha$-decay branch.'' This value is the currently adopted half-life.

\subsection*{$^{232}$Pu}
In the 1973 article ``The decay of the neutron deficient plutonium isotopes 232, 233, and 234'', J\"ager et al.\ identified $^{232}$Pu \cite{1973Jae01}. UO$_2$ targets, enriched in $^{233}$U were bombarded with 35 to 55~MeV alpha particles from the Karlsruhe Isochronous Cyclotron. Following chemical separation, $\alpha$- and $\gamma$-ray spectra were measured with a silicon surface-barrier and a Ge(Li) detector, respectively. ``The half-life of 34.1$\pm$0.7~minutes was obtained by the analysis of undisturbed $\alpha$-peaks originating from the members of the $\alpha$-decay family, e.g. $^{224}$Th, $^{220}$Ra, $^{212}$Po.'' This value agrees with the currently adopted half-life of 33.8(7)~min. The observation of $^{232}$Pu had previously been reported in an internal report \cite{1951Ort01}.

\subsection*{$^{233}$Pu}
Thomas et al.\ reported the discovery of $^{233}$Pu in the 1957 paper ``Decay properties of Pu$^{235}$, Pu$^{237}$, and a new isotope Pu$^{233}$ \cite{1957Tho01}. $^{233}$U targets were bombarded with 46~MeV $\alpha$ particles from the Berkeley 60-inch cyclotron forming $^{233}$Pu in the ($\alpha$,4n) reaction. Alpha-particle spectra were measured following chemical separation. ``The assignment of this activity to the previously unobserved nuclide Pu$^{233}$ is based primarily on three types of evidence: a rough excitation function, the appearance in the pulse analyses of alpha particles attributable to the U$^{229}$ daughter of Pu$^{233}$, and the compatibility of the alpha half-life with the alpha decay systematics.'' The measured half-life of 20(2)~min agrees with the presently adopted value of 20.9(4)~min.

\subsection*{$^{234}$Pu}
The discovery of $^{234}$Pu was described in the 1949 paper ``Products of the deuteron and helium-ion bombardments of U$^{233}$'' by Hyde et al.\ \cite{1949Hyd01}. $^{233}$U targets were bombarded with 44~MeV $\alpha$ particles from the Berkeley 60-in.\ cyclotron. Alpha-particle absorption and decay spectra were measured following chemical separation. ``By measuring the $\alpha$ spectrum of the plutonium fraction in a multichannel differential pulse analyzer at frequent intervals and plotting the specific decay of the 6.0-mev peak, values of 12.5 and 5.1~hr were obtained for the half life. The $\alpha$ half life may be taken as roughly 8$\pm$4~hr. This previously unknown plutonium isotope was tentatively identified as Pu$^{234}$ by establishing the presence of its U$^{230}$ daughter in the uranium fraction.'' The half-life is consistent with the currently adopted value of 8.8(1)~h.

\subsection*{$^{235}$Pu}
Thomas et al.\ reported the identification of $^{235}$Pu in the 1957 paper ``Decay properties of Pu$^{235}$, Pu$^{237}$, and a new isotope Pu$^{233}$ \cite{1957Tho01}. $^{233}$U and $^{235}$U targets were bombarded with $\alpha$ particles from the Berkeley 60-inch cyclotron forming $^{233}$Pu in the ($\alpha$,2n) and ($\alpha$,4n) reaction, respectively. Auger and conversion electrons from electron-capture decay were measured with a continous-flow-methane proportional counter following chemical separation. ``The values determined by Orth of 26$\pm$2~minutes for the over-all half-life and 5.85$\pm$0.03~Mev for the alpha energy have been confirmed by the present work.'' This value agrees with the currently adopted half-life of 25.3(5)~min. The previous work by Orth mentioned in the quote was an unpublished report \cite{1951Ort01}.

\subsection*{$^{236,237}$Pu}
James et al.\ identified $^{236}$Pu and $^{237}$Pu in  ``Products of helium-ion and deuteron bombardment of U$^{235}$ and U$^{238}$'' \cite{1949Jam01}. Natural uranium and $^{235}$U targets were bombarded with 16~MeV deuterons and 32~MeV $\alpha$ particles from the Berkeley 60-inch cyclotron. X rays, $\gamma$ rays and $\alpha$ particles were measured following chemical separation. ``Np$^{236}$ is a $\beta$-particle emitter with a half life of 22~hr. Its daughter, Pu$^{236}$, emits $\alpha$ particles with a range of 4.3~cm (energy 5.7~mev); it decays with a half life of 2.7~years. Np$^{235}$ decays by orbital-electron capture with a half life of approximately 400$\pm$20~days. A plutonium isotope that decays by orbital-electron capture with a half life of 40~days has been assigned the mass number 237.'' These half-lives of 2.7~y for $^{236}$Pu and 40~d for $^{237}$Pu agree with the currently adopted values of 2.858(8)~y and 45.64(4)~d, respectively.

\subsection*{$^{238}$Pu}
Kennedy et al.\ published the first confirmed observation of $^{238}$Pu in the 1949 paper ``Formation of the 50-year element 94 from deuteron bombardment of U$^{238}$'' \cite{1949Ken01}. An enriched $^{238}$U sample was bombarded with 14~MeV deuterons from the Berkeley 60-in.\ cyclotron. Following chemical separation, the growth of $\alpha$ particles was followed and aluminum absorption curves were measured with a Lauritsen electroscope. ``For these reasons, the most probable assignments of the 2.0-day element 93 and 50-year element 94 are 93$^{238}$ and 94$^{238}$ produced by the reactions U$^{238}$(d,2n)93$^{238}$, 93$^{238}\stackrel{\beta^-}{\longrightarrow}$ 2.0~day 94$^{238}$.'' This value is within a factor of two of the currently adopted half-life of 87.7(1)~y. The $^{238}$Pu activity had been observed earlier representing the discovery of the element $^{238}$Pu, however, no firm mass assignment was made \cite{1946Sea01}. The parent $\beta$-activity was assigned to either mass 238, 236 or 235.

\subsection*{$^{239}$Pu}
The discovery of $^{239}$Pu was reported by Kennedy et al.\ in the 1946 paper ``Properties of 94(239)'' \cite{1946Ken01}. A uranyl nitrate target was irradiated with neutrons produced by bombarding beryllium with 16~MeV deuterons from the Berkeley 60-inch cyclotron. $^{239}$Np was chemically separated and $^{239}$Pu was populated by $\beta$ decay. The sample was then irradiated with neutrons from the 37-inch cyclotron and subsequent fission events were observed with an ionization chamber.  A $^{239}$Pu sample was then placed near the screen window of an ionization chamber. ``After the 93$^{239}$ had decayed into 94$^{239}$ preliminary fission tests were made on this sample which then contained 0.5~microgram of 94$^{239}$. This sample was placed near the screen window of an ionization chamber which was imbedded in paraffin near the beryllium target of the 37-inch Berkeley cyclotron. This gave a small, but detectable, fission rate when a 6-microampere beam of deuterons was used.'' In addition, $\alpha$ decay was measured and a half-life of about 3$\times$10$^4$~years was extracted which is close to the currently adopted value of 24110(30)~y. It is interesting to note that the paper had been submitted already in 1941: ``This letter was received for publication on the date indicated [May 29, 1941] but was voluntarily withheld from publication until the end of the war. The original text has been somewhat changed, by omissions, in order to conform to present declassification standards.'' It should be noted that this paper submitted on May 29, 1941, was published in the October 1946 issue, while another paper by Kennedy and Wahl \cite{1946Ken02}, submitted on December 4, 1941, was published earlier in the April 1946 issue of Physical Review.

\subsection*{$^{240}$Pu}
James et al.\ identified $^{240}$Pu in ``Products of helium-ion and deuteron bombardment of U$^{235}$ and U$^{238}$'' \cite{1949Jam01}. Natural uranium and $^{238}$U targets were bombarded with 32~MeV $\alpha$ particles from the Berkeley 66-inch cyclotron forming $^{240}$Pu in the ($\alpha$,2n) reaction. X rays, $\gamma$ rays and $\alpha$ particles were measured following chemical separation. ``It was therefore concluded that this excess of $\alpha$ particles was due to Pu$^{240}$, which emits $\alpha$ particles with an energy very close to but slightly less than those of Pu$^{239}$. On the basis of this broadening, the range was taken as 3.60~cm of air; and from the yield of these $\alpha$ particles the half life of Pu$^{240}$ was estimated to be approximately 6,000~years.'' This value agrees with the currently adopted half-life of 6561(7)~y. James et al.\ credited a private communication by Chamberlain, Farwell and Segr\`e with the first observation of $^{240}$Pu.

\subsection*{$^{241}$Pu}
In 1949 Seaborg et al.\ reported the observation of $^{241}$Pu in the paper ``The new element americium (atomic number 95)'' \cite{1949Sea02}. 38~MeV $\alpha$ particles bombarded $^{238}$U targets from the Berkeley 60-in.\ cyclotron forming $^{241}$Pu in the $^{238}$U($\alpha$,n) reaction. Absorption curves were recorded and $\alpha$ and $\beta$ activities were measured following chemical separation. ``From these considerations the half-life of Pu$^{241}$ for $\beta$-particle emission is approximately 10~years.'' This value is close to the currently adopted half-life of 14.290(6)~y.

\subsection*{$^{242}$Pu}
$^{242}$Pu was discovered by Thompson et al.\ as described in the 1950 paper ``The new isotope Pu$^{242}$ and additional information on other plutonium isotopes'' \cite{1950Tho02}. $^{242}$Pu was produced by irradiating $^{239}$Pu and $^{241}$Am samples with neutrons. Following chemical separation, $^{242}$Pu was identified by mass spectroscopy. ``Alpha-pulse analysis of this plutonium showed the presence of alpha-particles of 4.88~Mev in abundance corresponding to a half-life of roughly 5$\times$10$^5$ years for Pu$^{242}$.'' This value is close to the currently adopted value of 3.75(2)$\times$10$^5$~y.

\subsection*{$^{243}$Pu}
Sullivan et al.\ identified $^{243}$Pu in the 1951 article ``Properties of plutonium-243'' \cite{1951Sul01}. A plutonium sample containing $^{239-242}$Pu was irradiated in the thimble of the Argonne heavy water reactor. Decay and absorptions curves were measured following chemical separation. ``From these results, we conclude that the five-hour activity is due to a plutonium isotopes, most probably Pu$^{243}$.'' This value agrees with the currently adopted half-life of 4.956(3)~h.

\subsection*{$^{244}$Pu}
In 1954 Studier et al.\ reported the observation of $^{244}$Pu in ``Plutonium-244 from pile-irradiated plutonium'' \cite{1954Stu01}. $^{239}$Pu was irradiated with neutrons in the Materials Testing Reactor. Plutonium was chemically separated and analyzed with a 12-inch, 60$^\circ$ mass spectrometer. ``Plutonium-244 is formed by Pu$^{243}$(n,$\gamma$)Pu$^{244}$ reaction and possibly by electron capture of Am$^{244}$.''

\subsection*{$^{245}$Pu}
$^{245}$Pu was discovered in 1955 simultaneously by Browne et al.\ in ``The decay chain Pu$^{245}-$Am$^{245}-$Cm$^{245}$'' \cite{1955Bro01} and Fields et al.\ in ``Production of Pu$^{245}$ and Am$^{245}$ by neutron irradiation of Pu$^{244}$'' \cite{1955Fie01}. Browne et al.\ produced $^{245}$Pu by neutron irradiation of transthorium elements. Beta- and gamma-ray spectra were measured following chemical separation. ``The half-life of Pu$^{245}$ was measured by resolving the gross decay curve of a plutonium sample, and by measuring the activity of its `equilibrated' daughter at successive times. The two methods lead to inconsistent values; direct decay gives a half-life of 12.8~hr, while methods based upon daughter activity lead to a value of 11.0~hr. Until further data are available, the best value to be taken is 12$\pm$1~hr.''
Fields et al.\ irradiated $^{244}$Pu with neutrons in the Argonne heavy water reactor. Absorption and decay curves were recorded and $\gamma$-ray spectra were measured with a sodium iodide crystal spectrometer. ``The beta-decay half-lives of Pu$^{245}$ and Am$^{245}$ were found to be 10.1$\pm$0.5 hours and 119$\pm$1 minutes, respectively.'' These half-lives for $^{245}$Pu agree with the currently adopted value of 10.5(1)~h.

\subsection*{$^{246}$Pu}
In the 1955 paper ``The new isotopes Pu$^{246}$ and Am$^{246}$'' Engelkemeir et al.\ reported the discovery of $^{246}$Pu \cite{1955Eng01}. $^{246}$Pu was detected in the debris of the November 1952 thermonuclear test. The mass assignment was made based on mass spectrometric measurements. Beta- and $\gamma$-ray spectra were measured with a scintillation spectrometer following chemical separation. ``The beta-emitting plutonium isotope decayed with a half-life of 11.2$\pm$0.2~days, while the half-life of the americium isotopes was 25$\pm$0.2~minutes... Re-examination of the Pu$^{246}$ content of this plutonium after 10 days disclosed a decrease in 246 mass consistent with a half-life of 11.2$\pm$0.2~days. On this basis the mass number was conclusively shown to be 246.'' The measured half-life of 11.2(2)~d for $^{246}$Pu agrees with the currently adopted half-life of 10.84(2)~d.

\subsection*{$^{247}$Pu}
The 1983 article ``Identification of $^{246}$Pu, $^{247}$Pu, $^{246m}$Am, and $^{247}$Am and determination of their half-lives'' by Popov et al.\ identified $^{247}$Pu \cite{1983Pop01}. Plutonium targets were irradiated with neutrons in a high-flux SM-2 reactor. X-ray, $\gamma$-ray, and $\alpha$-particle spectra were measured with a DGDK-30A Ge(Li)-detecting block, a BDRK-2-25 Si(Li) detector and a DKP s.d.-50 surface barrier Si(Au) detector, respectively, following chemical separation. In addition, the isotopic composition was measured with the MI 1201 mass spectrometer. ``As was noted above, in the plutonium fraction, after reaching equilibrium, a decrease in the intensity of the $\gamma$ lines 226 and 285~keV, belonging to $^{247}$Am, was observed. The curve of the decay of the nuclei of the parent substance is presented in [the figure]. The half-life, calculated according to these data, is 2.27$\pm$0.23~days. In our opinion, the parent nuclide is $^{247}$Pu, from which $^{247}$Am is formed according to the reaction $^{247}$Pu$\stackrel{\beta^-}{\longrightarrow}$2.27~days $^{247}$Am$\stackrel{\beta^-}{\longrightarrow}$23~min.'' This value is the currently adopted half-life.

\section{Discovery of $^{232-247}$Am}\vspace{0.0cm}

The element americium was discovered in 1944, announced in 1945 \cite{1945Sea01} and first published in 1949 by Seaborg et al.\ as part of the Plutonium Project Record identifying the isotopes $^{239}$Am, $^{240}$Am, and $^{241}$Am \cite{1949Sea02}. The name americium was officially accepted at the 15$^{th}$ IUPAC conference in Amsterdam in 1949 \cite{1949IUP01,2005Kop01}.

\begin{figure}
	\centering
	\includegraphics[scale=.7]{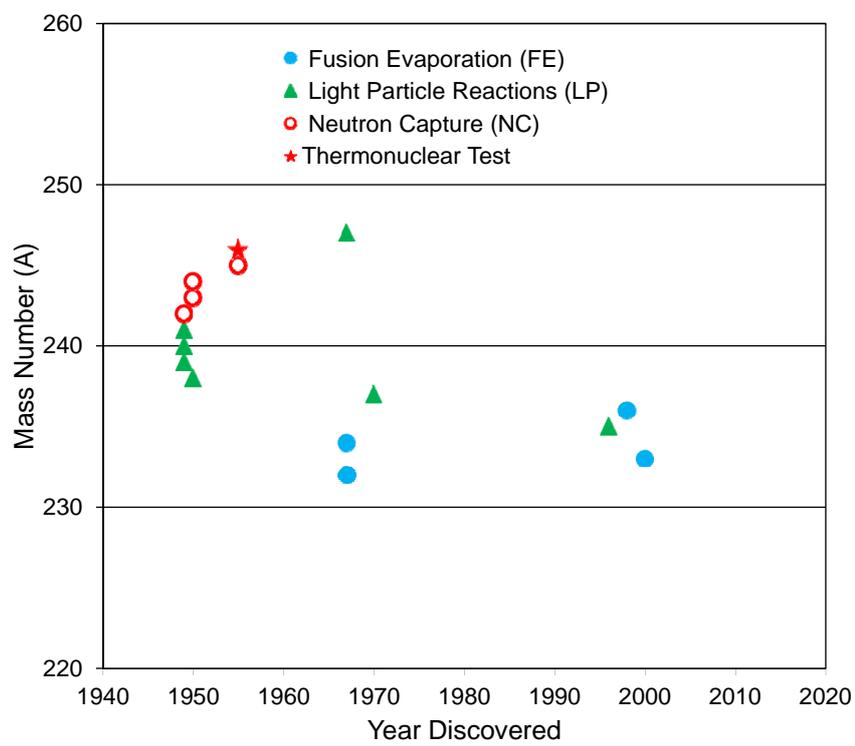}
	\caption{Americium isotopes as a function of time when they were discovered. The different production methods are indicated.}
\label{f:year-americium}
\end{figure}

Sixteen americium isotopes from A = 232--247 have been discovered so far. According to the HFB-14 model \cite{2007Gor01} about 80 additional americium isotopes could exist. Figure \ref{f:year-americium} summarizes the year of first discovery for all americium isotopes identified by the method of discovery: heavy-ion (A$>$4) fusion evaporation reactions (FE), light-particle (A$\le$4) reactions (LP), neutron capture (NC), and thermonuclear tests (TNT). In the following, the discovery of each americium isotope is discussed in detail and a summary is presented in Table 1. The observation of $^{230}$Am has only been reported in a progress report \cite{2003Mor01}.

\subsection*{$^{232}$Am}
Kuznetsov and Skobelev identified $^{232}$Am in the 1967 paper ``Investigation of 1.4-minute fissioning product in the Th$^{230}$ + B$^{10}$ reaction'' \cite{1967Kuz01}. $^{230}$Th was irradiated with a $^{10}$B beam forming $^{232}$Am in the (8n) fusion-evaporation reaction. The excitation function for spontaneous fission events was measured.  ``Estimates of the maximum of the excitation function half-width ($\sim$14~MeV) lead to the likely assumption that the 1.4-minute spontaneously-fissioning product is the result of the evaporation reaction Th$^{230}$(B$^{10}$,8n)$^{232}$Am,...'' The authors had reported this half-life previously without a mass assignment \cite{1967Kuz02}.

\subsection*{$^{233}$Am}
In the 2000 paper ``New isotope $^{233}$Am'' Sakama et al.\ reported the discovery of $^{233}$Am \cite{2000Sak01}. $^{233}$U was bombarded with a 63~MeV $^6$Li beam from the JAERI tandem accelerator producing $^{233}$Am in the (6n) fusion-evaporation reaction. $^{233}$Am was separated with the gas-jet coupled JAERI-ISOL on-line separator. Alpha-particles were measured with two Si PIN-photodiodes and X and $\gamma$ rays were measured with a HPGe detector. ``The $\alpha$-decay of $^{233}$Am and its subsequent $\alpha$-decay chain have been observed in the mass-233 fraction. The half-life and $\alpha$-particle energy of $^{233}$Am have been determined to be 3.2$\pm$0.8~min and 6780$\pm$17~keV, respectively.'' This value is the currently adopted half-life.

\subsection*{$^{234}$Am}
In 1967 Kuznetsov et al.\ identified $^{234}$Am in the paper ``Investigation of spontaneously fissile products in the reactions Th$^{230}$ + B$^{10}$ and Th$^{230}$ + B$^{11}$'' \cite{1967Kuz02}. Enriched $^{230}$Th targets were bombarded with $^{10}$B and $^{11}$B beams forming $^{234}$Am in the (6n) and (7n) reactions, respectively. Excitation functions were measured and spontaneous-fission fragments were detected. ``From this set of data it follows that the most probable product undergoing spontaneous fission T$_{1/2}$=2.6$\pm$0.2~min is Am$^{234}$.'' This value agrees with the currently adopted half-life of 2.32(8)~min. The authors had reported this half-life previously without a mass assignment \cite{1967Kuz04}.

\subsection*{$^{235}$Am}
1n 1996, Guo et al.\ announced the discovery of $^{235}$Am in ``A new neutron-deficient isotope, $^{235}$Am'' \cite{1996Guo01}. Enriched $^{238}$Pu targets were bombarded with 35~MeV protons from the Beijing proton linear accelerator producing $^{235}$Am in (p,4n) reactions. X- and $\gamma$-ray spectra were measured with a planar HPGe and a GMX HPGe detector following chemical separation. ``A radioactive series decay-analyzing program was used to fit the growth and decay curve for 101.1-keV Kx ray. Half-lives of 15$\pm$5~min for $^{235}$Am and 25$\pm$4~min for $^{235}$Pu were determined from the fit.'' This half-life of $^{235}$Am agrees with the currently adopted value of 9.9(5)~min.

\subsection*{$^{236}$Am}
$^{236}$Am was discovered by Tsukada et al.\ as described in the 1998 paper ``Half-life of the electron capture decaying isotope $^{236}$Am'' \cite{1998Tsu01}. Enriched $^{235}$U was bombarded with 46$-$50~MeV $^6$Li beams from the JAERI tandem accelerator producing $^{236}$Am in the (5n) fusion-evaporation reaction. $^{236}$Am was separated with the gas-jet coupled JAERI-ISOL on-line separator. X and $\gamma$ rays were measured with a planar type Ge and a coaxial n-type HPGe detector. ``From these facts, we conclude that the PuK X rays in [the figure] are solely ascribable to the EC decay of $^{236}$Am. The decay curves of the PuK$_{\alpha1}$ and K$_{\alpha2}$ X-ray intensities are shown in [the figure], and each half-life determined is 4.4$\pm$1.0 and 4.5$\pm$1.5~min, respectively. Thus the half-life of $^{236}$Am is evaluated to be 4.4$\pm$0.8 min.'' This value agrees with the currently adopted half-life of 3.6(2)~min. Previous evidence of a 0.6~y isomeric state of $^{236}$Am \cite{1987Mar01} has not been accepted by the ENSDF data evaluation.

\subsection*{$^{237}$Am}
In the 1970 paper ``Spontaneously fissioning isomers in U, Pu, Am and Cm isotopes'' Polikanov and Sletten identified $^{237}$Am \cite{1970Pol01}. A $^{238}$Pu target was bombarded with 12.0$-$14.1~MeV protons from the Copenhagen Van de Graaff accelerator. Fragments from fission-in-flight were measured in-flight with polycarbonate fission track detectors.``The 5~ns $^{237m}$Am fission isomer is observed and assigned through proton bombardments of $^{238}$Pu. The half-life is measured by the fission-in-flight method at 14.0~MeV proton energy.'' This value is the currently adopted half-life for the isomeric state. The ground state was first observed five years later \cite{1975Ahm01}.

\subsection*{$^{238}$Am}
Street et al.\ published ``The isotopes of americium'' in 1950, reporting the first observation of $^{238}$Am \cite{1950Str01}. $^{239}$Pu was bombarded with 50~MeV deuterons from the Berkeley 184-in. cyclotron. Differential counting with beryllium and lead absorbers was performed. ``In view of its half-life, radiation characteristics, and method of formation, this activity is probably an orbital electron capturing isotope and is best assigned to Am$^{238}$.'' The measured half-life of 1.2~h is close to the currently adopted half-life of 98(2)~min.

\subsection*{$^{239-241}$Am}
In 1949 Seaborg et al.\ reported the discovery of $^{239}$Am, $^{240}$Am, and $^{241}$Am in the paper ``The new element americium (atomic number 95)'' \cite{1949Sea02}. 38~MeV $\alpha$ particles bombarded $^{238}$U and $^{237}$Np targets and 19~MeV deuterons bombarded $^{239}$Pu targets from the Berkeley 60-in.\ cyclotron. $^{239}$Am was produced in the reactions $^{237}$Np($\alpha$,2n) and $^{239}$Pu(d,2n) and $^{240}$Am was produced in the reactions $^{237}$Np($\alpha$,n) and $^{239}$Pu(d,n). The reaction $^{238}$U($\alpha$,n) produced $^{241}$Pu which populated $^{241}$Am by $\beta$ decay. Absorption curves were recorded and $\alpha$ and $\beta$ activities were measured following chemical separation. ``Am$^{239}$, which undergoes branching decay, decaying (a) by orbital electron capture with a 12-hr half life and emitting 0.285-mev $\gamma$ rays and conversion electrons in addition to the characteristic X rays, and (b) by $\alpha$-particle emission (energy unknown) in the proportion of approximately 0.001 $\alpha$ particle per electron capture... Am$^{240}$, which decays by orbital-electron capture with a 50-hr half life, emitting 1.3- to 1.4-mev $\gamma$ rays and conversion electrons in addition to the characteristic X rays... This evidence proves that the $\alpha$ activity is due to Am$^{241}$ arising from the $\beta$-particle emission of Pu$^{241}$.'' The half-lives of 12-h for $^{239}$Am and 50-h for $^{240}$Am agree with the currently adopted values of 11.9(1)~h  and 50.8(3)~h, respectively. The observation of $^{241}$Am represented the discovery of the element americium. The half-life was later measured to be 510(20)~y \cite{1949Cun01}.

\subsection*{$^{242}$Am}
$^{242}$Am was identified by Manning and Asprey in 1949 in ``Preparation and radioactive properties of Am$^{242}$'' \cite{1949Man01}. $^{241}$Am was irradiated with neutrons in the thimble of the Argonne heavy-water pile. Alpha-particle spectra were measured following chemical separation. ``$_{95}$Am$^{241}$ captures neutrons to yield Am$^{242}$, a $\beta$ emitter with a half life of 16$\pm$3~hr. The maximum $\beta$ energy for Am$^{242}$ is approximately 1.0$\pm$0.3~mev.'' This half-life agrees with the presently adopted value of 16.02(2)~h. Seaborg et al.\ had produced $^{242}$Am earlier but did not measure any properties \cite{1949Sea02}. Seaborg et al.\ acknowledged the first half-life measurements of $^{242}$Am by Manning and Asprey.

\subsection*{$^{243,244}$Am}
Street et al.\ published ``The isotopes of americium'' in 1950, reporting the first observation of $^{243}$Am and $^{244}$Am \cite{1950Str01}. Americium was irradiated with neutrons in the Argonne uranium-heavy water pile forming $^{243}$Am in two successive neutron captures on $^{241}$Am and $^{244}$Am in the $^{243}$Am(n,$\gamma$) reaction. $^{243}$Am was identified by mass spectrometry and chemical analysis. No experimental details about $^{244}$Am were given. ``Mass spectrographic analysis of the americium of this bombardment showed Am$^{243}$ present to the extent of ca.\ 0.5 percent. This together with the yield of Np$^{239}$ determined in the chemical extraction experiments gives a partial half-life for alpha-particle emission for Am$^{243}$ of roughly 10$^4$~years... Irradiation of americium containing approximately ten percent of the isotope Am$^{243}$ with thermal neutrons in the uranium-heavy water pile at the Argonne Laboratory produced a new americium activity of ca.\ 25-min half-life at a yield corresponding to a cross section of roughly $\frac{1}{2}\times 10^{-22}$~cm$^2$. This activity is probably caused by the beta-emitting isotope Am$^{244}$, formed by an (n,$\gamma$) reaction.'' The $^{243}$Am half-life of 10$^4$~y is consistent with the currently adopted value of 7370(40)~y. The observed half-life of 25~min for $^{244}$Am corresponds to an isomeric state. The ground state was first observed twelve years later \cite{1962Van01}.

\subsection*{$^{245}$Am}
$^{245}$Am was discovered in 1955 simultaneously by Browne et al.\ in ``The decay chain Pu$^{245}-$Am$^{245}-$Cm$^{245}$'' \cite{1955Bro01} and Fields et al.\ in ``Production of Pu$^{245}$ and Am$^{245}$ by neutron irradiation of Pu$^{244}$'' \cite{1955Fie01}. Browne et al.\ produced $^{245}$Am by neutron irradiation of transthorium elements. Beta- and $\gamma$-ray spectra were measured following chemical separation. ``The 2.08-hour activity was observed to elute identically with the Am$^{241}$, which established its atomic number conclusively as 95... Accordingly, the first indication of the mass number of the new activity was obtained from the datum that more atoms of the parent were present than of the known Pu$^{246}$, leading to a tentative mass assignment of 245.'' Fields et al.\ irradiated $^{244}$Pu with neutrons in the Argonne heavy water reactor. Absorption and decay curves were recorded and $\gamma$-ray spectra were measured with a sodium iodide crystal spectrometer. ``The beta-decay half-lives of Pu$^{245}$ and Am$^{245}$ were found to be 10.1$\pm$0.5 hours and 119$\pm$1 minutes, respectively.'' These half-lives for $^{245}$Am agree with the currently adopted value of 2.05(1)~h.

\subsection*{$^{246}$Am}
In the 1955 paper ``The new isotopes Pu$^{246}$ and Am$^{246}$'' Engelkemeir et al.\ reported the discovery of $^{246}$Am \cite{1955Eng01}. $^{246}$Am was detected in the debris of the November 1952 thermonuclear test. The mass assignment was made based on mass spectrometric measurements. Beta- and $\gamma$-ray spectra were measured with a scintillation spectrometer following chemical separation. ``The beta-emitting plutonium isotope decayed with a half-life of 11.2$\pm$0.2~days, while the half-life of the americium isotopes was 25$\pm$0.2~minutes... Re-examination of the Pu$^{246}$ content of this plutonium after 10 days disclosed a decrease in 246 mass consistent with a half-life of 11.2$\pm$0.2~days. On this basis the mass number was conclusively shown to be 246.'' The measured half-life of 25.0(2)~min for $^{246}$Am corresponds to an isomeric state. The ground state was first observed twelve years later \cite{1967Ort01}.

\subsection*{$^{247}$Am}
In 1967, Orth et al.\ reported the discovery of $^{247}$Am in ``New short-lived americium beta emitters'' \cite{1967Ort01}. A $^{244}$Pu target was bombarded with 28~MeV $\alpha$ particles and 25~MeV $^3$He from the Los Alamos variable-energy cyclotron. Following chemical separation, $\gamma$-ray spectra were measured with a Ge(Li) detector. ``The new 24-min americium activity has been assigned to mass 247 on the following bases: (a) It was produced in $\alpha$-particle bombardment, but not in $^3$He bombardment of $^{244}$Pu. (b) The energy and M2 multipolarity of the 226-keV transition are consistent with the energy and 24-$\mu$sec half-life of the 226-keV state in $^{247}$Cm...'' This 24(3)~min half-life agrees with the currently adopted value of 23.0(13)~min.

\section{Discovery of $^{237-251}$Cm}\vspace{0.0cm}

The element curium was discovered in 1944, announced in 1945 \cite{1945Sea01} and first published in 1949 by Seaborg et al.\ as part of the Plutonium Project Record identifying the isotopes $^{240}$Cm and $^{242}$Cm \cite{1949Sea03}. The name curium was officially accepted at the 15$^{th}$ IUPAC conference in Amsterdam in 1949 \cite{1949IUP01,2005Kop01}.

\begin{figure}
	\centering
	\includegraphics[scale=.7]{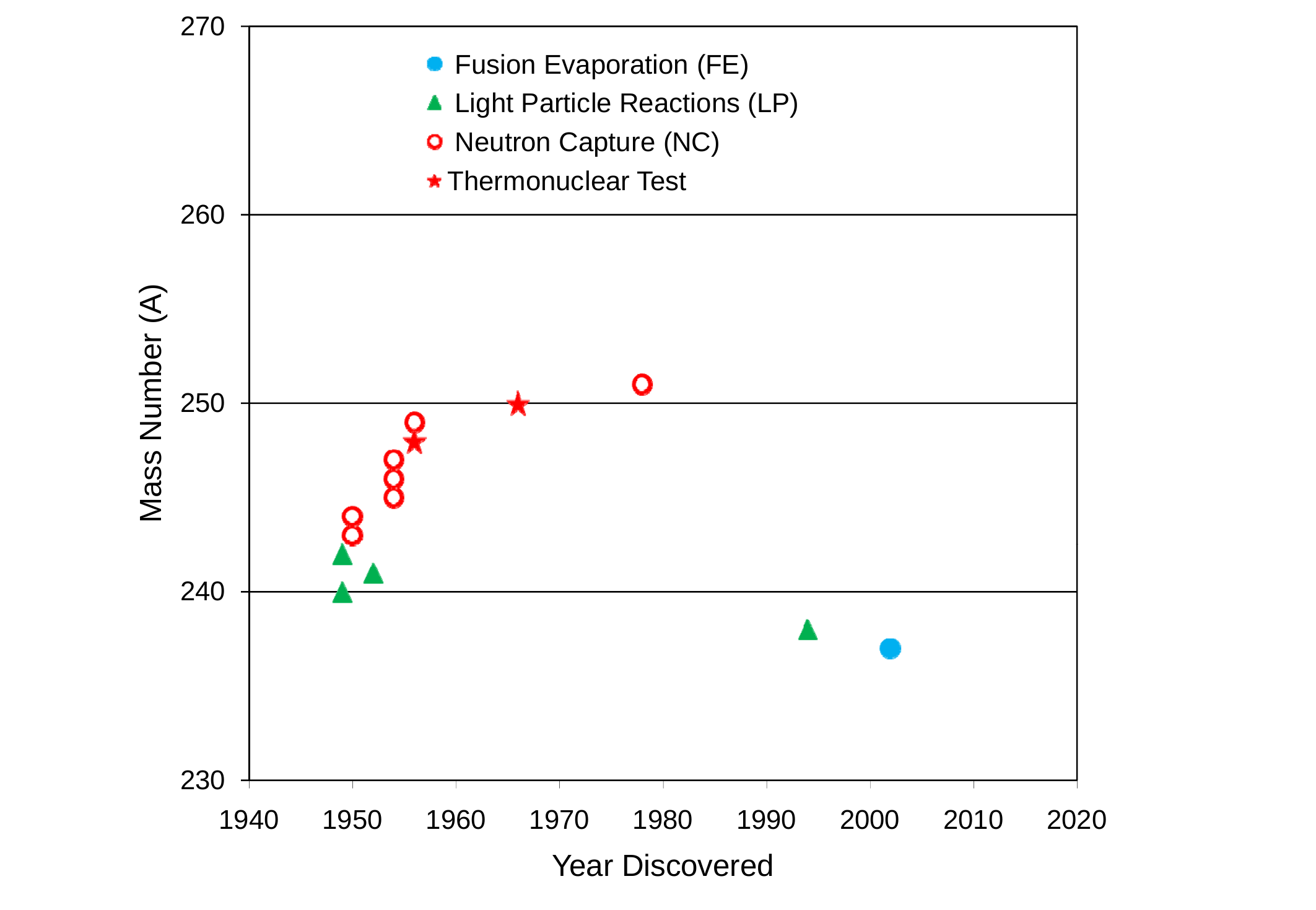}
	\caption{Curium isotopes as a function of time when they were discovered. The different production methods are indicated.}
\label{f:year-curium}
\end{figure}

Fourteen curium isotopes from A = 237--251 have been discovered so far. According to the HFB-14 model \cite{2007Gor01} about 85 additional curium isotopes could exist. Figure \ref{f:year-curium} summarizes the year of first discovery for all curium isotopes identified by the method of discovery: heavy-ion (A$>$4) fusion evaporation reactions (FE), light-particle (A$\le$4) reactions (LP), neutron capture (NC), and thermonuclear tests. In the following, the discovery of each curium isotope is discussed in detail and a summary is presented in Table 1.

In 1956 Glass et al.\ reported excitation functions for the $^{239}$Pu($\alpha$,4n)$^{239}$Cm without measuring any properties of $^{239}$Cm \cite{1956Gla01}. The 1967 Table of Isotopes \cite{1967Led01} listed half-lives of 3.0~h \cite{1952Car01} and 2.9~h \cite{1958Van01} which were based on private communications. These values are also quoted in the Nuclear Data Sheets for A=235,239 in 2003 \cite{2003Bro01}. Three $\alpha$-decay events assigned to $^{239}$Cm in an internal report \cite{2002Shi01} could not be confirmed \cite{2008Qin01}. Also, the observation of $^{233}$Cm and $^{234}$Cm has only been reported in an internal report \cite{2002Cag01}.

\subsection*{$^{237}$Cm}
Ichikawa et al.\ reported the discovery of $^{237}$Cm in the 2002 paper ``Performance of the multiple target He/PbI$_2$ aerosol jet system for mass separation of neutron-deficient actinide isotopes'' \cite{2002Ich01}. A stack of 21 $^{237}$Np targets were bombarded with a 62~MeV $^6$Li beam from the JAERI tandem accelerator forming $^{237}$Cm in the (6n) fusion-evaporation reaction.  $^{237}$Cm was separated with the on-line separator JAERI-ISOL and implanted in a Si detector which also recorded subsequent $\alpha$ decay. ``By comparing the $\alpha$ spectra [in the figures] the new $\alpha$-line with the energy of (6660$\pm$10)~keV observed at mass 237 fraction was assigned to $^{237}$Cm.''

\subsection*{$^{238}$Cm}
In the 1994 paper ``Electron-capture delayed fission properties of the new isotope $^{238}$Bk'' Kreek et al.\ reported the observation of $^{238}$Cm \cite{1994Kre02}. $^{241}$Am targets were bombarded with 75~MeV $\alpha$ particles from the Berkeley 88-inch cyclotron forming $^{238}$Bk in ($\alpha$,7n) reactions. $^{238}$Cm was then populated in $\beta$ decay. In addition, $\alpha$-particle spectra were measured following chemical separation. ``The alpha-decay chains for $^{238}$Bk and $^{238}$Cm are shown in [the figure]. The $^{238}$Cm decay was consistent with a 2.4-h half-life.'' The currently accepted half-life of 2.4(1)~h is based on a 1952 thesis \cite{1952Hig01}. In 1956 Glass et al.\ reported excitation functions for the reactions $^{239}$Pu($\alpha$,5n) and Pu$^{238}$($\alpha$,4n) without measuring any properties of $^{238}$Cm \cite{1956Gla01}.

\subsection*{$^{240}$Cm}
In 1949 Seaborg et al.\ reported the discovery of $^{240}$Cm in the paper ``The new element curium (atomic number 96)'' \cite{1949Sea03}. $^{240}$Cm was produced by bombarding plutonium targets with 40~MeV $\alpha$ particles from the Berkeley 60 in.\ cyclotron forming $^{240}$Cm in the reaction $^{239}$Pu($\alpha$,3n). Following chemical separation, $\alpha$ particles were measured in a parallel-plate ionization chamber. ``These isotopes are: (1) 96$^{242}$, which emits $\alpha$ particles with a range 4.75$\pm$0.1~cm in air and decays with a half life of 5.0$\pm$0.1~months; and (2) 96$^{240}$, which emits $\alpha$ particles with a range of 4.95$\pm$0.1~cm in air and decays with a half life of 26.8$\pm$0.3~days.'' This half-life for $^{240}$Cm agrees with the currently adopted value of 27(1)~d.

\subsection*{$^{241}$Cm}
$^{241}$Cm was identified in 1952 by Higgins and Street in ``The radiation characteristics of Cm$^{240}$ and Cm$^{241}$'' \cite{1952Hig02}. A $^{239}$Pu target was bombarded with 25$-$40~MeV $\alpha$ particles to form $^{241}$Cm in the ($\alpha$,2n) reaction. Conversion electrons and $\alpha$ particles were measured following chemical separation. ``When the bombardment energy was between 25 and 28 Mev, only the 35-day and 162-day activities were observed. At these energies the ($\alpha$,2n) reaction would be expected to predominate, while the ($\alpha$,3n) should be in such low incidence that products of it would not be detected. For this reason the 35-day activity was assigned to Cm$^{241}$.'' This half-life agrees with the currently adopted value of 32.8(2)~d. Seaborg et al.\ had earlier assigned a 55~d half-life to either $^{241}$Cm or $^{243}$Cm \cite{1949Sea03}.

\subsection*{$^{242}$Cm}
In 1949 Seaborg et al.\ reported the discovery of $^{242}$Cm in the paper ``The new element curium (atomic number 96)'' \cite{1949Sea03}. $^{242}$Cm was produced by bombarding plutonium targets with 32~MeV $\alpha$ particles from the Berkeley 60 in.\ cyclotron forming $^{242}$Cm in the reaction$^{239}$Pu($\alpha$,n). Following chemical separation, $\alpha$ particles were measured in a parallel-plate ionization chamber. ``These isotopes are: (1) 96$^{242}$, which emits $\alpha$ particles with a range 4.75$\pm$0.1~cm in air and decays with a half life of 5.0$\pm$0.1~months; and (2) 96$^{240}$, which emits $\alpha$ particles with a range of 4.95$\pm$0.1~cm in air and decays with a half life of 26.8$\pm$0.3~days.'' This half-life for $^{242}$Cm agrees with the currently adopted half-life of 162.8(2)~d.

\subsection*{$^{243,244}$Cm}
Reynolds et al.\ reported the observation of $^{243}$Cm and $^{244}$Cm in the 1950 article ``Mass-spectrographic identification of Cm$^{243}$ and Cm$^{244}$'' \cite{1950Rey02}. $^{243}$Cm and $^{244}$Cm were produced by neutron irradiation of $^{241}$Am and identified with a 60$^\circ$ focusing mass spectrograph following chemical separation. ``The isotopes Cm$^{243}$ and Cm$^{244}$ because of their small abundances are detected only at the more intense oxide masses 259 and 260.'' Earlier in the year Thompson et al.\ had reported tentative evidence for $^{243}$Cm but the calculated half-life of about 100~y \cite{1950Tho03} is significantly longer than the currently accepted value of 29.1(1)~y.

\subsection*{$^{245-247}$Cm}
In 1954 Stevens et al.\ identified $^{245}$Cm, $^{246}$Cm, and $^{247}$Cm in ``Curium isotopes 246 and 247 from pile-irradiated plutonium'' \cite{1954Ste01}. Plutonium samples were irradiated with neutrons in the Materials Testing Reactor. $^{245}$Cm, $^{246}$Cm, and $^{247}$Cm were identified with the Argonne 12-in.\ 60$^\circ$ mass spectrometer following chemically separation. ``Sample I contained 0.24 percent Cm$^{246}$, whereas sample II contained 1.27 percent Cm$^{246}$ and 0.016 percent Cm$^{247}$. Both curium samples also contained Cm$^{245}$, whose decay characteristics were recently identified.'' The previous identification of $^{245}$Cm mentioned in the quote referred to unpublished results by Hulet et al.\ and Reynolds \cite{1954Hul02}.

\subsection*{$^{248,249}$Cm}
The 1956 discovery of $^{248}$Cm and $^{249}$Cm by Fields et al.\ was reported in the paper ``Transplutonium elements in thermonuclear test debris'' \cite{1956Fie01}. $^{248}$Cm was detected in the debris of the 1952 thermonuclear test and $^{249}$Cm was produced by neutron irradiation of a curium fraction of the debris in the Materials Test Reactor. $^{248}$Cm and $^{249}$Cm were identified with the Argonne 12-in.\ 60$^\circ$ mass spectrometer following chemically separation. ``The curium was found to contain, in addition to the previously known Cm$^{245}$, the isotopes Cm$^{246}$, Cm$^{247}$, and Cm$^{248}$, in the mole percentages given in column (sample I) of [the table]... The isotope Cm$^{249}$ was made by MTR neutron bombardment of a curium fraction from the debris and found to have a half-life of 65~minutes and a beta energy of 0.9~Mev.'' The half-life measured for $^{249}$Cm agrees with the currently adopted value of 64.15(3)~min. Four months later Butler et al.\ independently reported a half-life 4.7(4)$\times$10$^5$~y for $^{248}$Cm \cite{1956But01}.

\subsection*{$^{250}$Cm}
In 1966 the Combined Radiochemistry Group reported the discovery of $^{250}$Cm in ``Nuclear decay properties of heavy nuclides produced in thermonuclear explosions-par and barbel events'' \cite{1966CRG01}. $^{250}$Cm was identified in the debris of the Par thermonuclear test. The number of $^{250}$Cm atoms was measured with a mass spectrometer following chemical separation. In addition, $\alpha$-decay and spontaneous fission events were recorded with ionization chambers. ``The resultant partial spontaneous-fission half-life for Cm$^{250}$ is (1.74$\pm$0.24)$\times$10$^4$~years.'' This value is within a factor of two of the currently adopted half-life of $\approx$8.3$\times$10$^3$~y. In 1956, Huizenga and Diamond estimated the $^{250}$Cm half-life from extrapolations of other curium abundances from the Mike thermonuclear test \cite{1956Hui01}.

\subsection*{$^{251}$Cm}
Lougheed et al.\ described the observation of $^{251}$Cm in the 1978 paper ``A new isotope of curium and its decay properties: $^{251}$Cm'' \cite{1978Lou01}. $^{250}$Cm from the Hutch thermonuclear explosion was irradiated with neutrons. Beta particles, X and $\gamma$ rays were measured following chemical separation. ``The weighted average of the half-life from all experiments is 16.8$\pm$0.2~min.'' This value is the currently accepted half-life.

\section{Discovery of $^{238-251}$Bk}\vspace{0.0cm}

The discovery of the element berkelium was published in 1950 by Thompson et al.\ identifying $^{243}$Bk \cite{1950Tho04}. The authors of this first paper suggested the name berkelium with the symbol Bk: ``It is suggested that element 97 be given the name berkelium (symbol Bk), after the city of Berkeley, in a manner similar to that used in naming its chemical homologue terbium (atomic number 65) whose name was derived from the town of Ytterby, Sweden, where the rare earth minerals were first found.''

\begin{figure}
	\centering
	\includegraphics[scale=.7]{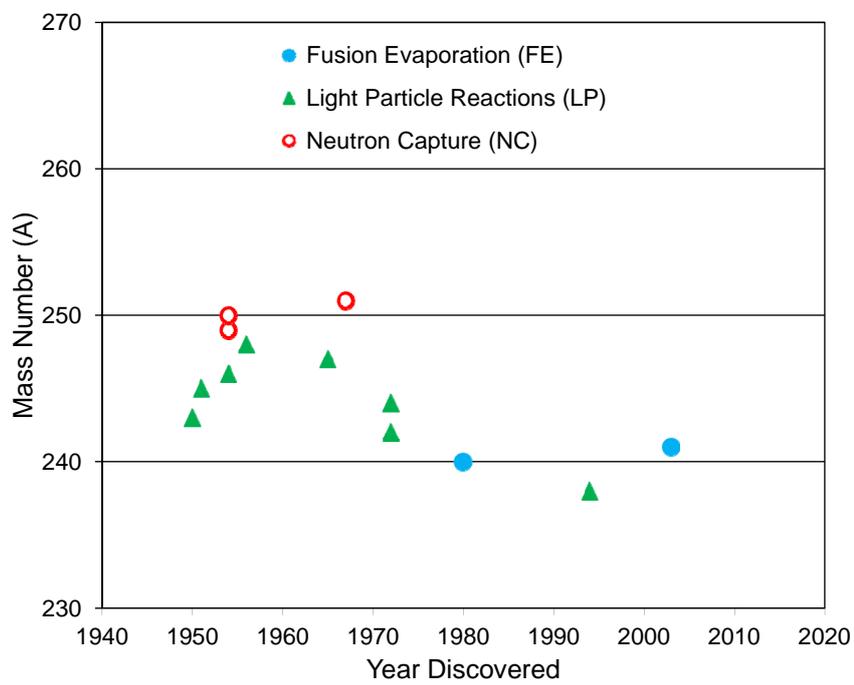}
	\caption{Berkelium isotopes as a function of time when they were discovered. The different production methods are indicated.}
\label{f:year-berkelium}
\end{figure}

Thirteen berkelium isotopes from A = 238--251 have been discovered so far. According to the HFB-14 model \cite{2007Gor01} about 80 additional berkelium isotopes could exist. Figure \ref{f:year-berkelium} summarizes the year of first discovery for all berkelium isotopes identified by the method of discovery: heavy-ion (A$>$4) fusion evaporation reactions (FE), light-particle (A$\le$4) reactions (LP), and neutron capture (NC). In the following, the discovery of each berkelium isotope is discussed in detail and a summary is presented in Table 1. The observation of $^{234}$Bk has only been reported in a progress report \cite{2003Mor01}. Evidence for the observation of $^{236}$Bk \cite{1987Mar01} has not been accepted by the ENSDF data evaluation.

\subsection*{$^{238}$Bk}
In the 1994 paper ``Electron-capture delayed fission properties of the new isotope $^{238}$Bk'' Kreek et al.\ reported the observation of $^{238}$Bk \cite{1994Kre02}. $^{241}$Am targets were bombarded with 75~MeV $\alpha$ particles from the Berkeley 88-inch cyclotron. X-ray-fission coincidences were measured with six pairs of passivated ion-implanted silicon detectors. In addition, $\alpha$-particle spectra were measured following chemical separation. ``Electron-capture delayed fission ECDF was studied in the new isotope $^{238}$Bk produced via the $^{241}$Am(75-MeV $\alpha$, 7n)$^{238}$Bk reaction. The half-life of the fission activity was measured to be 144$\pm$5 seconds.'' This value is the currently accepted half-life.

\subsection*{$^{240}$Bk}
In the 1980 article ``Study of delayed fission of the isotopes of Bk, Es, and Md'' Gangrskii et al.\ identified $^{240}$Bk \cite{1980Gan01}. A $^{232}$Th target was bombarded with a $^{14}$N beam from one of the Dubna cyclotrons forming $^{240}$Bk in the (6n) fusion-evaporation reaction. $^{240}$Bk was identified by electron-capture delayed fission where the fission fragments were measured with solid-state track detectors consisting of lavsan polyester films. ``The observed fission activities have half-lives close to those expected in fission after electron capture of $^{246}$Es (8~min) and $^{240}$Bk (the half-life 5$\pm$2~min obtained in the present work for this case is consistent with the value T$_{1/2}$ = 4~min which follows from the systematics for $^{240}$Bk).'' This half-life agrees with the currently adopted value of 4.8(8)~min.

\subsection*{$^{241}$Bk}
Asai et al.\ reported the discovery of $^{241}$Bk in the 2003 paper ``Identification of the new isotope $^{241}$Bk'' \cite{2003Asa01}. $^{239}$Pu targets were bombarded with 34-42~MeV $^6$Li beams from the JAERI tandem accelerator producing $^{241}$Bk in the (4n) fusion-evaporation reaction. $^{241}$Bk was separated with the gas-jet coupled JAERI-ISOL on-line separator. X and $\gamma$ rays were measured with a coaxial Ge detector and a 35\% n-type Ge detector. ``The half-lives of Cm K$_{\alpha 1}$ and K$_{\alpha 2}$ X rays were deduced through both the analyses. By taking a weighted average, the half-life of $^{241}$Bk was determined to be 4.6$\pm$0.4~min.'' This value is the currently adopted half-life.

\subsection*{$^{242}$Bk}
The identification of $^{242}$Bk by Wolf and Unik was reported in the 1972 paper ``Fissioning isomers of americium, curium and berkelium isotopes'' \cite{1972Wol01}. $^{241}$Am was bombarded with $\alpha$ particles from the Argonne 152~cm cyclotron forming $^{242}$Bk in the ($\alpha$,3n) reaction. $^{242}$Bk was identified from excitation functions of the delayed fission activity. ``[The figure] shows excitation functions obtained for the 9.5~ns and 600~ns half-life activities when $^{241}$Am was bombarded with $\alpha$ particles. The energy dependences of these excitation functions are unique for ($\alpha$,3n) reactions and indicate that both activities are due to isomeric states of $^{242}$Bk.'' These values are the accepted half-lives for the isomeric states. The ground state was first observed seven years later \cite{1979Wil01}.

\subsection*{$^{243}$Bk}
Thompson et al.\ reported the discovery of $^{243}$Bk in the 1950 paper ``Element 97'' \cite{1950Tho04}. A $^{241}$Am target was bombarded with $\alpha$ particles from the Berkeley 60-in. cyclotron. Alpha-particle spectra were measured following chemical separation. ``The particular isotope discovered is thought to be 97$^{243}$, or possibly 97$^{244}$, decaying with a 4.8-hour half-life by electron capture with approximately 0.1 percent alpha-decay branching.'' This value agrees with the currently adopted half-life of 4.5(2)~h. This measurement still represents the only half-life value for $^{243}$Bk published in the refereed literature. Two later measurements were only reported in Ph.D. theses \cite{1953Hul01,1956Che03}. Although this assignment was very tentative it is currently being accepted as correct \cite{2004Ako01}.

\subsection*{$^{244}$Bk}
The identification of $^{244}$Bk by Wolf and Unik was reported in the 1972 paper ``Fissioning isomers of americium, curium and berkelium isotopes'' \cite{1972Wol01}. $^{243}$Am was bombarded with $\alpha$ particles from the Argonne 152~cm cyclotron forming $^{244}$Bk in the ($\alpha$,3n) reaction. $^{244}$Bk was identified from excitation functions of the delayed fission activity. ``A fissioning isomer with an 820~ns half-life has been assigned to $^{244}$Bk on the basis of the measured characteristic ($\alpha$,3n) excitation function shown in [the figure].'' This value is the currently adopted half-life for the isomeric state. Previously, the observation of the $^{244}$Bk ground-state was reported in two unpublished theses \cite{1956Che03,1966Ahm01}.

\subsection*{$^{245}$Bk}
Hulet et al.\ identified $^{245}$Bk in the 1951 paper  ``New isotopes of berkelium and californium'' \cite{1951Hul01}. A curium target was bombarded with 35~MeV $\alpha$ particles and 16~MeV deuterons from the Berkeley 60-in.\ cyclotron. Alpha-particle spectra were measured in a windowless gas proportional counter following chemical separation. ``Consideration of the systematics of alpha-radioactivity suggests that the new 4.95-day isotope is most likely Bk$^{245}$.'' This value is included in the calculation to obtain the currently adopted half-life.

\subsection*{$^{246}$Bk}
In the 1954 paper ``Isotopes of curium, berkelium, and californium'' Hulet et al.\ reported the discovery of $^{246}$Bk \cite{1954Hul01}. Americium targets were bombarded with 27~MeV $\alpha$-particles from the Berkeley 60-in.\ cyclotron. Following chemical separation X and $\gamma$ rays were measured with a sodium iodide crystal scintillation counter and decay curves were recorded with a windowless proportional counter. ``The bombardment of various mixtures of Am$^{241}$ and Am$^{243}$ with 27-Mev helium ions is believed to have produced a 1.8-day electron-capture isotope of berkelium which is tentatively assigned to Bk$^{246}$.'' This value agrees with the currently adopted half-life of 1.80(2)~d.

\subsection*{$^{247}$Bk}
In 1965 Milsted et al.\ reported the discovery of $^{247}$Bk in the paper ``The alpha half-life of berkelium-247; A new long-lived isomer of berkelium-248'' \cite{1965Mil01}. A curium target was bombarded with $\alpha$ particles from the Argonne 150~cm cyclotron. The $\alpha$-decay half-life of $^{247}$Bk was measured by a mass-spectrometric isotopic dilution method. ``The 247/249 mass ratio, taken in conjunction with the activity ratio, gives an alpha half-life of 1380$\pm$250~y for Bk$^{247}$, if the $\beta^-$ half-life of Bk$^{240}$ is taken 310~d.'' This value is the currently adopted half-life. The observation of $^{247}$Bk had previously been reported in a Ph.\ D.\ thesis \cite{1956Che03}.

\subsection*{$^{248}$Bk}
The first observation of $^{248}$Bk was described in the 1956 article ``New isotope of berkelium'' by Hulet \cite{1956Hul01}. A curium target was bombarded with 25~MeV $\alpha$ particles from the Berkeley 60-in.\ cyclotron. $^{248}$Bk was identified by periodically separating $^{248}$Cf and measuring its $\alpha$-decay. ``A 23$\pm$5~hour half-life was calculated for Bk$^{248}$ from the amount of Cf$^{248}$ grown, the time intervals associated with the growths and decays, and estimated chemical yields.'' This half-life agrees with the value of 23.7(2)~h for an isomeric state.

\subsection*{$^{249}$Bk}
The 1954 discovery of $^{249}$Bk was reported in ``Identification of californium isotopes 249, 250, 251, and 252 from pile-irradiated plutonium'' by Diamond et al.\ \cite{1954Dia01}. Plutonium was irradiated with neutrons in the Idaho Materials Testing Reactor. Activities were measured following chemical separation. ``The alpha half-life of Cf$^{249}$ is calculated to be 550$\pm$150 years from the 5.81-Mev alpha-particle growth rate into a Bk$^{249}$ sample of a measured disintegration rate.'' In a table the half-life of $^{249}$Bk is listed as $\sim$1~y which is consistent with the presently accepted value of 330(4)~d. Previously, Thompson et al.\ had tentatively assigned a $\beta$-activity to $^{249}$Bk with a lower half-life limit of one week \cite{1954Tho01}.

\subsection*{$^{250}$Bk}
In 1954 Ghiorso et al.\ reported the discovery of $^{250}$Bk in the article ``New isotopes of americium, berkelium and californium'' \cite{1954Ghi02}. $^{250}$Bk was produced by neutron capture on $^{249}$Bk. Alpha particles were measured following chemical separation. ``A sample of Bk$^{249}$ was subjected to a short neutron bombardment, followed by chemical purification, and a new activity, presumably Bk$^{250}$, was produced. It decayed by $\beta^-$ emission with a half-life of 3.13~hours.'' This value agrees with the currently adopted half-life of 3.212(5)~h.

\subsection*{$^{251}$Bk}
Diamond et al.\ identified $^{251}$Bk in the 1967 paper ``Nuclear properties of $^{251}$Bk'' \cite{1967Dia01}. $^{255}$Es was produced by neutron irradiation of a pure einsteinium sample and $^{251}$Bk was populated by $\alpha$ decay. $^{255}$Es was separated with the Argonne Isotope Separator and the $\alpha$ recoils were collected on collodion films. The subsequent $\beta$ decay of $^{251}$Bk was measured with an end-window proportional counter.  ``A new isotope, $^{251}$Bk, has been isolated and found to decay by $\beta^-$ particle emission with a half-life of 57.0$\pm$1.7~min.'' This value agrees with the currently adopted half-life of 55.6(11)~min.

\section{Discovery of $^{237-256}$Cf}\vspace{0.0cm}

The discovery of the element californium was published in 1950 by Thompson et al. \cite{1950Tho07}. The authors assigned a 45~min half-life tentatively to $^{244}$Cf, however, this assignment was later changed to $^{245}$Cf \cite{1956Che02}. Thus, the original paper by Thompson et al.\ constitutes the discovery of the element californium but not a californium isotope. The authors of this first paper suggested the name californium with the symbol Cf: ``It is suggested that element 98 be given the name californium (symbol Cf) after the university and state where the work was done. This name, chosen for the reason given, does not reflect the observed chemical homology of element 98 to dysprosium (No. 66) as the names americium (No. 95), curium (No. 96), and berkelium (No. 97) signify that these elements are the chemical homologs of europium (No. 63), gadolinium (No. 64), and terbium (No. 65), respectively; the best we can do is point out,
in recognition of the fact that dysprosium is named on the basis of a Greek word meaning `difficult to get at,' that the searchers for another element a century ago found it difficult to get to California'' \cite{1950Tho07}.

\begin{figure}
	\centering
	\includegraphics[scale=.7]{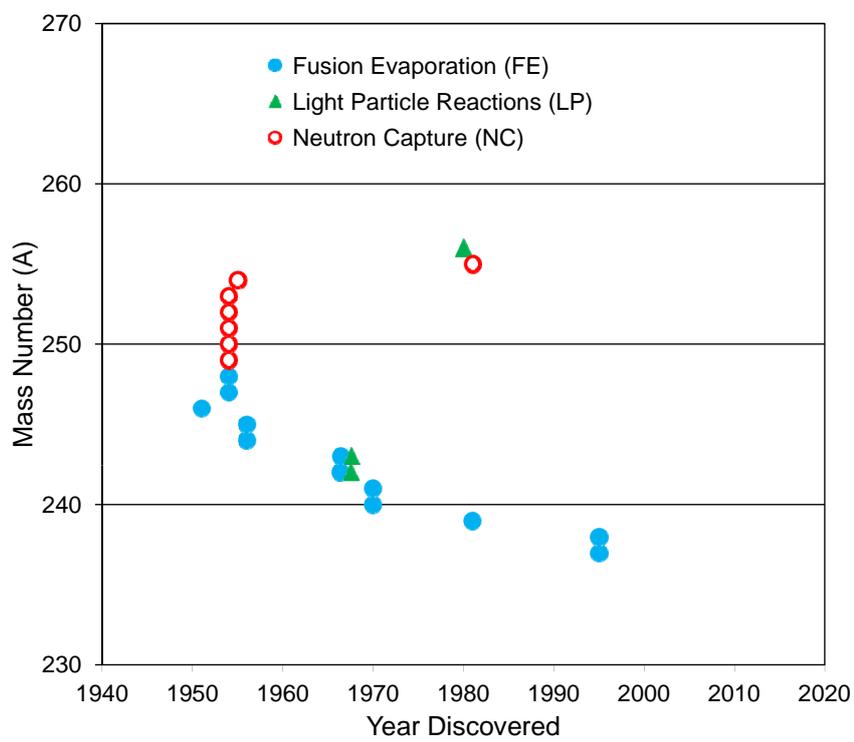}
	\caption{Californium isotopes as a function of time when they were discovered. The different production methods are indicated.}
\label{f:year-californium}
\end{figure}

Twenty californium isotopes from A = 237--256 have been discovered so far. According to the HFB-14 model \cite{2007Gor01} about 85 additional californium isotopes could exist. Figure \ref{f:year-californium} summarizes the year of first discovery for all californium isotopes identified by the method of discovery: heavy-ion (A$>$4) fusion evaporation reactions (FE), light-particle (A$\le$4) reactions (LP), and neutron capture (NC). In the following, the discovery of each californium isotope is discussed in detail and a summary is presented in Table 1.

\subsection*{$^{237,238}$Cf}
$^{237}$Cf and $^{238}$Cf were discovered in 1995 by Lazarev et al.\ and the results were reported in the paper ``Spontaneous fission of light californium isotopes produced in $^{206,207,208}$Pb $+$ $^{34,36}$S reactions; new nuclide $^{238}$Cf'' \cite{1995Laz01}. Enriched $^{206}$Pb, $^{207}$Pb and $^{208}$Pb targets were bombarded with 215~MeV $^{34}$S and $^{36}$S beams from the Dubna U400 cyclotron. $^{237}$Cf was formed in the fusion evaporation reactions $^{206}$Pb($^{34}$S,4n) and $^{207}$Pb($^{34}$S,3n) and $^{238}$Cf was formed with $^{34}$S on $^{206,207,208}$Pb and $^{36}$S on $^{206}$Pb. Mica fission-fragment detectors arranged around a rotating target cylinder detected spontaneous fission events. ``We identified a new spontaneously fissioning isotope $^{238}$Cf with T$_{sf}\approx$T$_{1/2}$=21$\pm$2~ms and obtained evidence of the production of a new isotope $^{237}$Cf with T$_{1/2}$=2.1$\pm$0.3~s.'' These half-lives correspond to the currently adopted values for $^{237}$Cf and $^{238}$Cf.

\subsection*{$^{239}$Cf}
M\"unzenberg et al.\ reported the discovery of $^{239}$Cf in 1981 in ``The new isotopes $^{247}$Md, $^{243}$Fm,$^{239}$Cf, and investigation of the evaporation residues from fusion of $^{206}$Pb, $^{208}$Pb, and $^{209}$Bi with $^{40}$Ar'' \cite{1981Mun01}. A $^{206}$Pb target was bombarded with a 4.8~MeV/u $^{40}$Ar beam from the GSI UNILAC accelerator to form $^{243}$Fm in the (3n) fusion-evaporation reaction. Recoil products were separated with the velocity filter SHIP and implanted in an array of position sensitive surface-barrier detectors which also recorded subsequent $\alpha$ decay and spontaneous fission. ``Correlated to this decay we observed a daughter decay of (7,630$\pm$25)keV and a half1ife of (39$^{+37}_{-12}$)s. We assign these two decays to $^{243}$Fm and its daughter $^{239}$Cf.'' This value is the currently adopted half-life.

\subsection*{$^{240,241}$Cf}
The 1970 discovery of $^{240}$Cf and $^{241}$Cf by Silva et al.\ was described in the paper ``New isotopes $^{241}$Cf and $^{240}$Cf'' \cite{1970Sil01}. Enriched $^{233}$U, $^{234}$U, and $^{235}$U targets were bombarded with 118~MeV $^{12}$C beams from the Oak Ridge isochronous cyclotron. Reaction products were deposited with a helium jet on a platinum disk which was then moved in front of a Si(Au) detector to measure $\alpha$-particles. ``The half-lives of the 7.335- and 7.590-MeV $\alpha$  activities were determined to be 3.78$\pm$0.70 and 1.06$\pm$0.15~min, respectively, from computer least squares fits to 200-300 recorded decay events... the excitation functions for the 7.335- and 7.590-MeV activities support the assignment of the former $\alpha$-particle activity to the decay of $^{241}$Cf and of the latter to $^{240}$Cf.'' The $^{241}$Cf half-life is the currently adopted value and the $^{240}$Cf half-life is included in the calculation of the average value of the presently accepted half-life. Previously a 7.31~MeV peak had tentatively been assigned to $^{241}$Cf \cite{1967Fie01}, however Silva et al.\ reassigned this peak to a state of $^{242}$Cf.

\subsection*{$^{242}$Cf}
$^{242}$Cf was simultaneously discovered in 1967 by Sikkeland and Ghiorso in ``New californium isotope, $^{242}$Cf'' \cite{1967Sik01} and Fields et al.\ in ``Nuclear properties of $^{242}$Cf, $^{243}$Cf, $^{244}$Cf, and $^{245}$Cf'' \cite{1967Fie01}. Sikkeland and Ghiorso used the Berkeley Hilac to bombard uranium targets with a 124~MeV $^{12}$C beam. Reaction products were slowed in helium gas and deposited on a platinum disk which was then moved to an $\alpha$ grid chamber. ``A least-squares-fit analysis of the decays for the $\alpha$ group at 7.39$\pm$0.02~MeV in which about 500 events were used gave a half-life of 3.4$\pm$0.2~min. The shapes and positions of the maxima of excitation functions for the production of this $\alpha$ emitter corresponded to a ($^{12}$C,3n), ($^{12}$C,4n), ($^{12}$C,5n), ($^{12}$C,6n) and ($^{12}$C,8n) reactions with the targets $^{233}$U, $^{234}$U, $^{235}$U, $^{236}$U and $^{238}$U, respectively, and is thus the nuclide $^{242}$Cf.'' Fields et al.\ bombarded $^{242}$Cm and $^{244}$Cm targets with a $^3$He beam from the Argonne 60-in.\ cyclotron. The subsequent $\alpha$ decay of the recoils was measured. ``Irradiations of $^{242}$Cm and $^{244}$Cm by $^3$He ions produced two new isotopes of californium; $^{242}$Cf emits a 7.35$\pm$0.01~MeV $\alpha$-particle group and has a half-life of 3.2$\pm$0.5~min. $^{243}$Cf emits 7.06$\pm$0.01 and 7.17$\pm$0.01~MeV $\alpha$-particle groups and decays with a half-life of 12.5$\pm$1.0~min.'' These values agree with the currently adopted half-lives of 3.7(5)~min.

\subsection*{$^{243}$Cf}
$^{243}$Cf was simultaneously discovered in 1967 by Sikkeland et al.\ in ``Decay properties of the new isotope $^{243}$Cf and of $^{244}$Cf'' \cite{1967Sik02} and Fields et al.\ in ``Nuclear properties of $^{242}$Cf, $^{243}$Cf, $^{244}$Cf, and $^{245}$Cf'' \cite{1967Fie01}. Sikkeland et al.\ used the Berkeley Hilac to bombard uranium targets with a 124~MeV $^{12}$C beam. Reaction products were slowed in helium gas and deposited on a platinum disk which was then moved to an $\alpha$ grid chamber. ``A least-squares analysis of the decay of the $\alpha$-particle group at 7.05$\pm$0.02~MeV in which about 300 events were used gave a half-life of 10.3$\pm$0.5~min. The assignment of the emitter to $^{243}$Cf was based on the excitation functions.'' Fields et al.\ bombarded $^{242}$Cm and $^{244}$Cm targets with a $^3$He beam from the Argonne 60-in.\ cyclotron. The subsequent $\alpha$ decay of the recoils was measured. ``Irradiations of $^{242}$Cm and $^{244}$Cm by $^3$He ions produced two new isotopes of californium; $^{242}$Cf emits a 7.35$\pm$0.01~MeV $\alpha$-particle group and has a half-life of 3.2$\pm$0.5~min. $^{243}$Cf emits 7.06$\pm$0.01 and 7.17$\pm$0.01~MeV $\alpha$-particle groups and decays with a half-life of 12.5$\pm$1.0~min.'' These values agree with the currently adopted half-life of 10.7(5)~min.

\subsection*{$^{244,245}$Cf}
In the 1956 article ``Mass assignment of the 44-minute californium-245 and the new isotope californium-244'', Chetham-Strode et al.\ identified $^{244}$Cf and $^{245}$Cf \cite{1956Che02}. Samples of $^{244}$Cm were bombarded with $\alpha$ particles from the Berkeley 60-in.\ cyclotron. Subsequent emission of $\alpha$ particles was measured following chemical separation. ``The 44-minute californium alpha emitter previously thought to be Cf$^{244}$ has been reassigned to mass number 245 on the basis of milking experiments, excitation functions, cross bombardments, and decay scheme studies. Californium-245 decays by the emission of (7.11$\pm$0.02)-Mev alpha particles ($\sim$30\%) and by orbital electron capture ($\sim$70\%). The new isotope Cf$^{244}$ was also identified and found to decay by the emission of (7.1$\pm$0.02)-Mev alpha particles with a half-life of 25$\pm$3 minutes.'' These values are close to the currently adopted half-lives of 19.4(6)~min and 45.0(15)~min for $^{244}$Cf and $^{245}$Cf, respectively. In the paper reporting the discovery of the element californium the 45~min half-life had tentatively been assigned to $^{244}$Cf instead of $^{245}$Cf \cite{1950Tho07,1950Tho06}.

\subsection*{$^{246}$Cf}
$^{246}$Cf was discovered by Ghiorso et al.\ as described in the 1951 article ``Californium isotopes from bombardment of uranium with carbon ions'' \cite{1951Ghi02}. Uranium metal targets were bombarded with a 120~MeV $^{12}$C beam from the Berkeley 60-inch cyclotron. Alpha-particle spectra were measured with an ionization chamber following chemical separation. ``A consideration of the systematics of alpha-radioactivity leads us to the view that this 35-hour period is due to the new isotope Cf$^{246}$ formed by the reaction U$^{238}$(C$^{12}$,4n).'' The quoted value agrees with the currently adopted half-life of 35.7(5)~h.

\subsection*{$^{247,248}$Cf}
Ghiorso et al. discovered $^{247}$Cf and $^{248}$Cf in 1954 as reported in ``Reactions of U$^{238}$ with cyclotron-produced nitrogen ions'' \cite{1954Ghi01}. The Berkeley 60-inch cyclotron was used to bombard a $^{238}$U target with $^{14}$N beams. Resulting activities were measured following chemical separation. No further details were given and the results were summarized in a table. ``The following transmutation products have been observed: 99$^{247(?)}$, 99$^{246}$, Cf$^{244}$, Cf$^{246}$, Cf$^{247(?)}$, Cf$^{248}$, Bk$^{243}$, and other berkelium isotopes not yet identified.'' The measured half-lives of $\sim$2.7~hr for Cf$^{247}$ and 225~days for $^{248}$Cf are close to the currently adopted values of 3.11(3)~h and 333.5(28)~d, respectively.

\subsection*{$^{249,250}$Cf}
In 1954 Ghiorso et al.\ reported the discovery of $^{249}$Cf and $^{250}$Cf in the article ``New isotopes of americium, berkelium and californium'' \cite{1954Ghi02}. $^{249}$Bk was produced by neutron irradiation of $^{239}$Pu in the Materials Testing Reactor. {$^{249}$Cf was then populated in the decay of $^{249}$Bk while $^{250}$Cf was formed in decay of $^{250}$Bk which was produced by further neutron irradiation of $^{249}$Bk. The resulting activities were measured following chemical separation. ``From the alpha disintegration rate and the beta disintegration rate of the Bk$^{250}$ parent, the alpha half-life of Cf$^{250}$ was found to be about 12~years... From the amount of Cf$^{249}$ alpha activity which grew from a known amount of Bk$^{249}$, the alpha half-life of the Cf$^{249}$ was found to be about 400~years.'' These values are close to the currently adopted half-lives of 351(2)~y and 13.08(9)~y for $^{249}$Cf and $^{250}$Cf, respectively. Less than a week later Diamond et al.\ independently reported half-lives of 550(150)~y and 9.4(32)~y for $^{249}$Cf and $^{250}$Cf, respectively \cite{1954Dia01}.

\subsection*{$^{251}$Cf}
The 1954 discovery of $^{251}$Cf was reported in ``Identification of californium isotopes 249, 250, 251, and 252 from pile-irradiated plutonium'' by Diamond et al.\ \cite{1954Dia01}. Plutonium was irradiated with neutrons in the Idaho Materials Testing Reactor. $^{251}$Cf was identified with a 12-inch, 60$^\circ$ mass spectrometer following chemical separation. ``Californium isotopes of mass numbers 249, 250, 251, and 252 were detected in mole percentages given in column 1 of [the table].''

\subsection*{$^{252}$Cf}
In 1954 Ghiorso et al.\ reported the discovery of $^{252}$Cf in the article ``New isotopes of americium, berkelium and californium'' \cite{1954Ghi02}. $^{249}$Bk was irradiated with neutrons producing $^{252}$Cf. Alpha particles were measured following chemical separation. ``Direct alpha-decay measurements performed over a period of several months on a sample consisting largely of Cf$^{252}$ indicate a half-life for this nuclide of roughly two years.'' This value is close to the currently adopted half-life of 2.645(8)~y. Less than a week later Diamond et al.\ independently reported a half-life of 2.1(4)~y \cite{1954Dia01}.

\subsection*{$^{253}$Cf}
Choppin et al. reported the first observation of $^{253}$Cf in 1954 in their paper ``Nuclear properties of some isotopes of californium, elements 99 and 100'' \cite{1954Cho01}. $^{253}$Cf was produced by neutron irradiation of $^{239}$Pu in the Materials Testing Reactor. Resulting activities were measured following chemical separation. ``The nuclide Cf$^{253}$ decays by $\beta^-$ emission to 99$^{253}$. From the rate of growth of the 99$^{253}$ alpha activity in the purified Cf$^{253}$ sample and the rate of decay of separated 99$^{253}$, the half-life of both these nuclides was found to be approximately twenty days.'' This half-life is consistent with the presently adopted value of 17.81(8)~d. Less than a week later Diamond et al.\ independently reported a half-life of 18(3)~d \cite{1954Dia01}.

\subsection*{$^{254}$Cf}
The 1955 article ``Nuclide 99$^{254}$'' described the first observation of $^{254}$Cf by Harvey et al.\ \cite{1955Har01}. $^{254}$Cf was produced by neutron irradiation of $^{239}$Pu in the Materials Testing Reactor. Resulting activities were measured following chemical separation. ``Preliminary experiments showed that a californium isotope decaying by spontaneous fission, with no detectable emission of
alpha particles, grew into very carefully purified samples containing 99$^{253}$, 99$^{254}$, 99$^{254m}$, and 99$^{255}$. The californium exhibited a half-life of 85$\pm$15~days... The californium isotope most responsible for such a short-lived spontaneous fission decay is most likely even mass, and is therefore probably Cf$^{254}$.'' This value is close to the currently adopted half-life of 60.5(2)~d.

\subsection*{$^{255}$Cf}
In 1981, Lougheed et al.\ reported the observation of $^{255}$Cf in ``Two new isotopes with N = 157: $^{256}$Es and $^{255}$Cf'' \cite{1981Lou01}. A $^{254}$Cf sample from the Hutch underground nuclear explosion was irradiated with neutrons in the General Electric Vallecitos Test Reactor. Alpha-particle spectra were measured following chemical separation. ``For $^{255}$Cf, we determined a half-life of 85$\pm$18 (2$\sigma$)~min from the weighted average of the two experiments.'' This half-life is the currently adopted value.

\subsection*{$^{256}$Cf}
Hoffman et al.\ reported the discovery of $^{256}$Cf in the 1980 paper ``12.3-min $^{256}$Cf and 43-min $^{258}$Md and systematics of the spontaneous fission properties of heavy nuclides'' \cite{1980Hof01}. A $^{254}$Cf target was bombarded with a 16~MeV triton beam from the Los Alamos Van de Graaff accelerator forming $^{256}$Cf in the reaction $^{254}$Cf(t,p). Recoil products were collected on carbon foils and then moved in front of three pairs of Si(Au) surface-barrier detectors to measure fragments from spontaneous fission. ``A total of 766 binary coincidence events was recorded in on-line scalers associated with the three detector pairs. A least-mean-squares fit to this counting data is consistent with the presence of two components, one with a half-life of 12.3$\pm$1.2~min, and one long-lived species that we attribute to $^{254}$Cf.'' This value is the currently adopted half-life.

\section{Summary}
The discoveries of the known transuranium isotopes neptunium, plutonium, americium, curium, berkelium, and californium have been compiled and the methods of their production discussed. The identification of these isotopes was relatively straightforward. The half-lives of $^{228,234}$Np, $^{232,234}$Am, and $^{241}$Cm were initially reported with no or only uncertain mass assignments and the first half-life reported for $^{243}$Cm was incorrect. In addition, an $\alpha$-decay energy originally assigned to $^{241}$Cf was later reassigned to $^{242}$Cf. Also, in the californium discovery paper the assignment of the observed half-life to $^{244}$Cf was later changed to $^{245}$Cf.

\ack

This work was supported by the National Science Foundation under grants No. PHY06-06007 (NSCL) and PHY10-62410 (REU).

\bibliography{../isotope-discovery-references}

\newpage

\newpage

\TableExplanation

\bigskip
\renewcommand{\arraystretch}{1.0}

\section{Table 1.\label{tbl1te} Discovery of neptunium, plutonium, americium, curium, berkelium, and californium isotopes }
\begin{tabular*}{0.95\textwidth}{@{}@{\extracolsep{\fill}}lp{5.5in}@{}}
\multicolumn{2}{p{0.95\textwidth}}{ }\\

Isotope & Neptunium, plutonium, americium, curium, berkelium, and californium  isotope \\
Author & First author of refereed publication \\
Journal & Journal of publication \\
Ref. & Reference \\
Method & Production method used in the discovery: \\

  & FE: heavy-ion fusion evaporation \\
  & LP: light-particle reactions (including neutrons) \\
  & NC: neutron capture reactions \\
  & TR: heavy-ion transfer reactions \\
  & TNT: thermonuclear tests \\

Laboratory & Laboratory where the experiment was performed\\
Country & Country of laboratory\\
Year & Year of discovery \\
\end{tabular*}
\label{tableI}

\datatables 



\setlength{\LTleft}{0pt}
\setlength{\LTright}{0pt}


\setlength{\tabcolsep}{0.5\tabcolsep}

\renewcommand{\arraystretch}{1.0}

\footnotesize 

\begin{longtable}{@{\extracolsep\fill}llllllll@{}}
\caption{Discovery of neptunium, plutonium, americium, curium, berkelium, and californium isotopes. See page\ \pageref{tbl1te} for Explanation of Tables}
Isotope & Author & Journal & Ref. & Method & Laboratory & Country & Year\\
\hline\\
\endfirsthead\\
\caption[]{(continued)}
Isotope & Author & Journal & Ref. & Method & Laboratory & Country & Year\\
\hline\\
\endhead
$^{225}$Np & A.V. Yeremin & Nucl. Instrum. Meth. A &\cite{1994Yer01}& FE & Dubna & Russia &1994 \\
$^{226}$Np & V. Ninov & Z. Phys. A &\cite{1990Nin01}& FE & Darmstadt & Germany &1990 \\
$^{227}$Np & A.N. Andreyev & Z. Phys. A &\cite{1990And01}& FE & Dubna & Russia &1990 \\
$^{228}$Np & S.A. Kreek & Phys. Rev. C &\cite{1994Kre01}& LP & Berkeley & USA &1994 \\
$^{229}$Np & R.L. Hahn & Nucl. Phys. A &\cite{1968Hah01}& LP & Oak Ridge & USA &1968 \\
$^{230}$Np & R.L. Hahn & Nucl. Phys. A &\cite{1968Hah01}& LP & Oak Ridge & USA &1968 \\
$^{231}$Np & L.B. Magnusson & Phys. Rev. &\cite{1950Mag01}& LP & Berkeley & USA &1950 \\
$^{232}$Np & L.B. Magnusson & Phys. Rev. &\cite{1950Mag01}& LP & Berkeley & USA &1950 \\
$^{233}$Np & L.B. Magnusson & Phys. Rev. &\cite{1950Mag01}& LP & Berkeley & USA &1950 \\
$^{234}$Np & E.K. Hyde & Nat. Nucl. Ener. Ser. &\cite{1949Hyd01}& LP & Berkeley & USA &1949 \\
$^{235}$Np & R.A. James & Nat. Nucl. Ener. Ser. &\cite{1949Jam01}& LP & Berkeley & USA &1949 \\
$^{236}$Np & R.A. James & Nat. Nucl. Ener. Ser. &\cite{1949Jam01}& LP & Berkeley & USA &1949 \\
$^{237}$Np & A.C. Wahl & Phys. Rev. &\cite{1948Wah01}& LP & Berkeley & USA &1948 \\
$^{238}$Np & J.W. Kennedy & Nat. Nucl. Ener. Ser. &\cite{1949Ken01}& LP & Berkeley & USA &1949 \\
$^{239}$Np & E. McMillan & Phys. Rev. &\cite{1940McM01}& LP & Berkeley & USA &1940 \\
$^{240}$Np & J.D. Knight & Phys. Rev. &\cite{1953Kni01}& NC & Los Alamos & USA &1953 \\
$^{241}$Np & R. Vandenbosch & Phys. Rev. &\cite{1959Van01}& LP & Argonne & USA &1959 \\
$^{242}$Np & P.E. Haustein & Phys. Rev. C &\cite{1979Hau01}& LP & Brookhaven & USA &1979 \\
$^{243}$Np & E.R. Flynn & Phys. Rev. C &\cite{1979Fly01}& LP & Los Alamos & USA &1979 \\
$^{244}$Np & K.J. Moody & Z. Phys. A &\cite{1987Moo01}& TR & Darmstadt & Germany &1987 \\
 & & & & & & & \\
 & & & & & & & \\
$^{228}$Pu & A.N. Andreyev & Z. Phys. A &\cite{1994And01}& FE & Dubna & Russia &1994 \\
$^{229}$Pu & A.N. Andreyev & Z. Phys. A &\cite{1994And01}& FE & Dubna & Russia &1994 \\
$^{230}$Pu & A.N. Andreyev & Z. Phys. A &\cite{1990And03}& FE & Dubna & Russia &1990 \\
$^{231}$Pu & C.A. Laue & Phys. Rev. C &\cite{1999Lau01}& LP & Berkeley & USA &1999 \\
$^{232}$Pu & U. J\"ager& Z. Phys. &\cite{1973Jae01}& LP & Karlsruhe & Germany &1973 \\
$^{233}$Pu & T.D. Thomas & Phys. Rev. &\cite{1957Tho01}& LP & Berkeley & USA &1957 \\
$^{234}$Pu & E.K. Hyde & Nat. Nucl. Ener. Ser. &\cite{1949Hyd01}& LP & Berkeley & USA &1949 \\
$^{235}$Pu & T.D. Thomas & Phys. Rev. &\cite{1957Tho01}& LP & Berkeley & USA &1957 \\
$^{236}$Pu & R.A. James & Nat. Nucl. Ener. Ser. &\cite{1949Jam01}& LP & Berkeley & USA &1949 \\
$^{237}$Pu & R.A. James & Nat. Nucl. Ener. Ser. &\cite{1949Jam01}& LP & Berkeley & USA &1949 \\
$^{238}$Pu & J.W. Kennedy & Nat. Nucl. Ener. Ser. &\cite{1949Ken01}& LP & Berkeley & USA &1949 \\
$^{239}$Pu & J.W. Kennedy & Phys. Rev. &\cite{1946Ken01}& LP & Berkeley & USA &1946 \\
$^{240}$Pu & R.A. James & Nat. Nucl. Ener. Ser. &\cite{1949Jam01}& LP & Berkeley & USA &1949 \\
$^{241}$Pu & G.T. Seaborg & Nat. Nucl. Ener. Ser. &\cite{1949Sea02}& LP & Berkeley & USA &1949 \\
$^{242}$Pu & S.G. Thompson & Phys. Rev. &\cite{1950Tho02}& NC & Berkeley & USA &1950 \\
$^{243}$Pu & J.C. Sullivan & Phys. Rev. &\cite{1951Sul01}& NC & Argonne & USA &1951 \\
$^{244}$Pu & M.H. Studier & Phys. Rev. &\cite{1954Stu01}& NC & Argonne & USA &1954 \\
$^{245}$Pu & C.I. Browne & J. Inorg. Nucl. Chem. &\cite{1955Bro01}& NC & Los Alamos & USA &1955 \\
 & P.R. Fields & J. Inorg. Nucl. Chem. &\cite{1955Fie01}& NC & Argonne & USA &1955 \\
$^{246}$Pu & D. Engelkemeir & J. Inorg. Nucl. Chem. &\cite{1955Eng01}& TNT& Argonne & USA &1955 \\
$^{247}$Pu & Yu.S. Popov & Sov. Radiochem. &\cite{1983Pop01}& NC & Dimitrovgrad & Russia &1983 \\
 & & & & & & & \\
 & & & & & & & \\
$^{232}$Am & V.I. Kuznetsov & Sov. J. Nucl. Phys. &\cite{1967Kuz01}& FE & Dubna & Russia &1967 \\
$^{233}$Am & M. Sakama & Eur. Phys. J. A &\cite{2000Sak01}& FE & JAERI & Japan &2000 \\
$^{234}$Am & V.I. Kuznetsov & Sov. J. Nucl. Phys. &\cite{1967Kuz02}& FE & Dubna & Russia &1967 \\
$^{235}$Am & J. Guo & Z. Phys. A &\cite{1996Guo01}& LP & Beijing & China &1996 \\
$^{236}$Am & K. Tsukada & Phys. Rev. C &\cite{1998Tsu01}& FE & JAERI & Japan &1998 \\
$^{237}$Am & S.M. Polikanov & Nucl. Phys. A &\cite{1970Pol01}& LP & Copenhagen & Denmark &1970 \\
$^{238}$Am & K. Street & Phys. Rev. &\cite{1950Str01}& LP & Berkeley & USA &1950 \\
$^{239}$Am & G.T. Seaborg & Nat. Nucl. Ener. Ser. &\cite{1949Sea02}& LP & Berkeley & USA &1949 \\
$^{240}$Am & G.T. Seaborg & Nat. Nucl. Ener. Ser. &\cite{1949Sea02}& LP & Berkeley & USA &1949 \\
$^{241}$Am & G.T. Seaborg & Nat. Nucl. Ener. Ser. &\cite{1949Sea02}& LP & Berkeley & USA &1949 \\
$^{242}$Am & W. M. Manning & Nat. Nucl. Ener. Ser. &\cite{1949Man01}& NC & Argonne & USA &1949 \\
$^{243}$Am & K. Street & Phys. Rev. &\cite{1950Str01}& NC & Argonne & USA &1950 \\
$^{244}$Am & K. Street & Phys. Rev. &\cite{1950Str01}& NC & Argonne & USA &1950 \\
$^{245}$Am & C.I. Browne & J. Inorg. Nucl. Chem. &\cite{1955Bro01}& NC & Los Alamos & USA &1955 \\
 & P.R. Fields & J. Inorg. Nucl. Chem. &\cite{1955Fie01}& NC & Argonne & USA &1955 \\
$^{246}$Am & D. Engelkemeir & J. Inorg. Nucl. Chem. &\cite{1955Eng01}& TNT& Argonne & USA &1955 \\
$^{247}$Am & C.J. Orth & Phys. Rev. Lett. &\cite{1967Ort01}& LP & Los Alamos & USA &1967 \\
 & & & & & & & \\
 & & & & & & & \\
$^{237}$Cm & S. Ichikawa & Nucl. Instrum. Meth. B &\cite{2002Ich01}& FE & JAERI & Japan &2002 \\
$^{238}$Cm & S.A. Kreek & Phys. Rev. C &\cite{1994Kre02}& LP & Berkeley & USA &1994 \\
$^{239}$Cm & & & & & & & \\
$^{240}$Cm & G.T. Seaborg & Nat. Nucl. Ener. Ser. &\cite{1949Sea03}& LP & Berkeley & USA &1949 \\
$^{241}$Cm & G.H. Higgins & Phys. Rev. &\cite{1952Hig02}& LP & Berkeley & USA &1952 \\
$^{242}$Cm & G.T. Seaborg & Nat. Nucl. Ener. Ser. &\cite{1949Sea03}& LP & Berkeley & USA &1949 \\
$^{243}$Cm & F.L. Reynolds& Phys. Rev. &\cite{1950Rey02}& NC & Berkeley & USA &1950 \\
$^{244}$Cm & F.L. Reynolds& Phys. Rev. &\cite{1950Rey02}& NC & Berkeley & USA &1950 \\
$^{245}$Cm & C.M. Stevens & Phys. Rev. &\cite{1954Ste01}& NC & Argonne & USA &1954 \\
$^{246}$Cm & C.M. Stevens & Phys. Rev. &\cite{1954Ste01}& NC & Argonne & USA &1954 \\
$^{247}$Cm & C.M. Stevens & Phys. Rev. &\cite{1954Ste01}& NC & Argonne & USA &1954 \\
$^{248}$Cm & P.R. Fields & Phys. Rev. &\cite{1956Fie01}& TNT& Argonne & USA &1956 \\
$^{249}$Cm & P.R. Fields & Phys. Rev. &\cite{1956Fie01}& NC & Argonne & USA &1956 \\
$^{250}$Cm & Combined Radiochemistry Group & Phys. Rev. &\cite{1966CRG01}& TNT& Livermore & USA &1966 \\
$^{251}$Cm & R.W. Lougheed & J. Inorg. Nucl. Chem. &\cite{1978Lou01}& NC & Livermore & USA &1978 \\
 & & & & & & & \\
 & & & & & & & \\
$^{238}$Bk & S.A. Kreek & Phys. Rev. C &\cite{1994Kre02}& LP & Berkeley & USA &1994 \\
$^{239}$Bk & & & & & & & \\
$^{240}$Bk & Yu.P. Gangrskii & Sov. J. Nucl. Phys. &\cite{1980Gan01}& FE & Dubna & Russia &1980 \\
$^{241}$Bk & M. Asai & Eur. Phys. J. A &\cite{2003Asa01}& FE & JAERI & Japan &2003 \\
$^{242}$Bk & K.L. Wolf & Phys. Lett. B &\cite{1972Wol01}& LP & Argonne & USA &1972 \\
$^{243}$Bk & S.G. Thompson & Phys. Rev. &\cite{1950Tho04}& LP & Berkeley & USA &1950 \\
$^{244}$Bk & K.L. Wolf & Phys. Lett. B &\cite{1972Wol01}& LP & Argonne & USA &1972 \\
$^{245}$Bk & E.K. Hulet & Phys. Rev. &\cite{1951Hul01}& LP & Berkeley & USA &1951 \\
$^{246}$Bk & E.K. Hulet & Phys. Rev. &\cite{1954Hul01}& LP & Berkeley & USA &1954 \\
$^{247}$Bk & J. Milsted & Nucl. Phys. &\cite{1965Mil01}& LP & Argonne & USA &1965 \\
$^{248}$Bk & E.K. Hulet & Phys. Rev. &\cite{1956Hul01}& LP & Berkeley & USA &1956 \\
$^{249}$Bk & H. Diamond & Phys. Rev. &\cite{1954Dia01}& NC & Argonne & USA &1954 \\
$^{250}$Bk & A. Ghiorso & Phys. Rev. &\cite{1954Ghi02}& NC & Berkeley & USA &1954 \\
$^{251}$Bk & H. Diamond & J. Inorg. Nucl. Chem. &\cite{1967Dia01}& NC & Argonne & USA &1967 \\
 & & & & & & & \\
 & & & & & & & \\
$^{237}$Cf & Yu.A. Lazarev & Nucl. Phys. A &\cite{1995Laz01}& FE & Dubna & Russia &1995 \\
$^{238}$Cf & Yu.A. Lazarev & Nucl. Phys. A &\cite{1995Laz01}& FE & Dubna & Russia &1995 \\
$^{239}$Cf & G. M\"unzenberg & Z. Phys. A &\cite{1981Mun01}& FE & Darmstadt & Germany &1981 \\
$^{240}$Cf & R.J. Silva & Phys. Rev. C &\cite{1970Sil01}& FE & Oak Ridge & USA &1970 \\
$^{241}$Cf & R.J. Silva & Phys. Rev. C &\cite{1970Sil01}& FE & Oak Ridge & USA &1970 \\
$^{242}$Cf & T. Sikkeland & Phys. Lett. B &\cite{1967Sik01}& FE & Berkeley & USA &1967 \\
 & P.R. Fields & Phys. Lett. B &\cite{1967Fie01}& LP & Argonne & USA &1967 \\
$^{243}$Cf & T. Sikkeland & Phys. Lett. B &\cite{1967Sik02}& FE & Berkeley & USA &1967 \\
 & P.R. Fields & Phys. Lett. B &\cite{1967Fie01}& LP & Argonne & USA &1967 \\
$^{244}$Cf & A. Chetham-Strode & Phys. Rev. &\cite{1956Che02}& FE & Berkeley & USA &1956 \\
$^{245}$Cf & A. Chetham-Strode & Phys. Rev. &\cite{1956Che02}& FE & Berkeley & USA &1956 \\
$^{246}$Cf & A. Ghiorso & Phys. Rev. &\cite{1951Ghi02}& FE & Berkeley & USA &1951 \\
$^{247}$Cf & A. Ghiorso & Phys. Rev. &\cite{1954Ghi01}& FE & Berkeley & USA &1954 \\
$^{248}$Cf & A. Ghiorso & Phys. Rev. &\cite{1954Ghi01}& FE & Berkeley & USA &1954 \\
$^{249}$Cf & A. Ghiorso & Phys. Rev. &\cite{1954Ghi02}& NC & Berkeley & USA &1954 \\
$^{250}$Cf & A. Ghiorso & Phys. Rev. &\cite{1954Ghi02}& NC & Berkeley & USA &1954 \\
$^{251}$Cf & H. Diamond & Phys. Rev. &\cite{1954Dia01}& NC & Argonne & USA &1954 \\
$^{252}$Cf & A. Ghiorso & Phys. Rev. &\cite{1954Ghi02}& NC & Berkeley & USA &1954 \\
$^{253}$Cf & G.R. Choppin & Phys. Rev. &\cite{1954Cho01}& NC & Berkeley & USA &1954 \\
$^{254}$Cf & B.G. Harvey & Phys. Rev. &\cite{1955Har01}& NC & Berkeley & USA &1955 \\
$^{255}$Cf & R.W. Lougheed & J. Inorg. Nucl. Chem. &\cite{1981Lou01}& NC & Livermore & USA &1981 \\
$^{256}$Cf & D.C. Hoffman & Phys. Rev. C &\cite{1980Hof01}& LP & Los Alamos & USA &1980 \\
 \\
\end{longtable}

\end{document}